\documentclass[aps,prb,longbibliography
]{revtex4-2}

\usepackage{graphicx}% Include figure files
\usepackage{dcolumn}% Align table columns on decimal point
\usepackage{bm}% bold math
\usepackage{amsmath}
\usepackage{cancel}
\usepackage{float}
\usepackage{ifsym}
\usepackage{color}
\usepackage{latexsym}
\usepackage{latexsym}
\usepackage{dcolumn}
\usepackage{epsfig}
\usepackage{amsfonts}
\usepackage{amsmath}
\usepackage{dsfont}
\usepackage{accents}
\usepackage[all,cmtip]{xy}
\usepackage{paralist}
\usepackage{enumerate}
\usepackage{wasysym}
\usepackage{hyperref}
\usepackage{enumitem}
\usepackage{multirow}
\usepackage{caption,subcaption}

\begin{document}
%\linenumbers
%\preprint{}

\title{Symmetry and polarity of antiphase boundaries in PbZrO$_3$} %Title of paper

\author{I.~Rychetsky}
\email{rychet@fzu.cz}
\affiliation{Institute of Physics of the Czech Academy of Sciences, Na Slovance 2, 18221 Prague 8, Czech Repbublic.}
\author{W.~Schranz}
%\email{wilfried.schranz@univie.ac.at}
\affiliation{University of Vienna, Faculty of Physics, Boltzmanngasse 5, 1090 Wien, Austria.}
\author{A.~Tr\"{o}ster}
\affiliation{University of Vienna, Faculty of Physics, Boltzmanngasse 5, 1090 Wien, Austria.}

\date{\today}

\begin{abstract}
The polar properties of antiphase boundaries (APBs) in PbZrO$_3$ are analyzed in detail using a recently developed layer group approach in order parameter space and compared with the results from  Landau-Ginzburg free energy description. It is shown that the former approach reveals the microscopic APBs' properties, and predicts polar APB structures at particular positions inside the unit cell, which agree very well with recent experimental obsevations [Wei, et.~al. \cite{Wei2014,Wei2015}]. The systematic usage of the method is developed. In contrast with it the commonly used free energy description obscures the microscopic features but still can reflect the macroscopic properties of the APBs by taking into account the bilinear coupling of polarization and order parameter gradients. The relation between the layer group approach and the Landau-Ginzburg free energy description is discussed and two mechanisms of polarization switching inside the APBs are distinguished. It is illustrated that the polar APBs observed in PbZrO$_3$ are consistently and naturally explained by the layer group approach. This analysis is expected to have a significant impact also in other materials.
\end{abstract}

 %Grain and twin boundaries, Domain structure 
\maketitle

%\begin{widetext}
\section{Introduction}
	Domain walls (DWs) in ferroic materials became recently of increasing interest due to the substantial improvement of experimental techniques allowing their observation \cite{Aert2012,Yokota2014,Yokota2017,Salje2013} and unveiling their potential for applications \cite{Lei2018,Nataf2020,Catalan2012} . The tensor properties of DWs and in particular of antiphase boundaries (APBs) in perovskites were studied by several authors \cite{Salje2013,Salje2016,Salje2020,Gu2014,Stengel2017,Wei2014}. The occurrence of polarization inside the DWs in otherwise non-polar samples was predicted on the basis of symmetry analysis using the layer-group method \cite{Janovec2006,Janovec1989} and also described by the  phenomenological Landau-Ginzburg approach \cite{Ishibashi1976,Bullbich1989,Salje1991,Tagantsev2001,Sonin1989,Wei2014}. In the latter case it is explained by flexoelectricity occurring at the DWs' center due to the coupling of polarization and the strain gradient \cite{Morozovska2012,Gu2014,Tagantsev2013}. It turned out that besides flexoelectricity the so-called 'roto-polar' effect can also be responsible for polarization, which is driven by a coupling of polarization and a Lifshitz-like gradient term of the order parameter (OP) \cite{Stengel2017,Salje1991}. A general approach based on the combination of OP symmetry and layer group analysis was suggested and demonstrated for the study of DWs' properties in KSCN \cite{Schranz2019}, $\mathrm{SrTiO_3}$ (STO) \cite{Schranz2020}, and $\mathrm{PbZrO_3}$ (PZO) \cite{Schranz2020(2)}. The observation and modeling of the microscopic polar structure of APBs in antiferroelectric PZO were presented in Refs. \cite{Wei2014,Wei2015,Ma2019}, and the appearance of polarization at the APB center was explained by a  biquadratic coupling between OP and polarization \cite{Wei2014}. 
	
	In the paper we present an analysis of the APBs' polar properties in PZO, especially linking between microscopic and macroscopic properties, and a manifestation of the microscopic symmetry in the phenomenological Landau-Ginzburg description. 
	  
\section{Symmetry and domain states in lead zirconate}
	
\begin{figure} 
		\centering
		\includegraphics[scale=0.5]{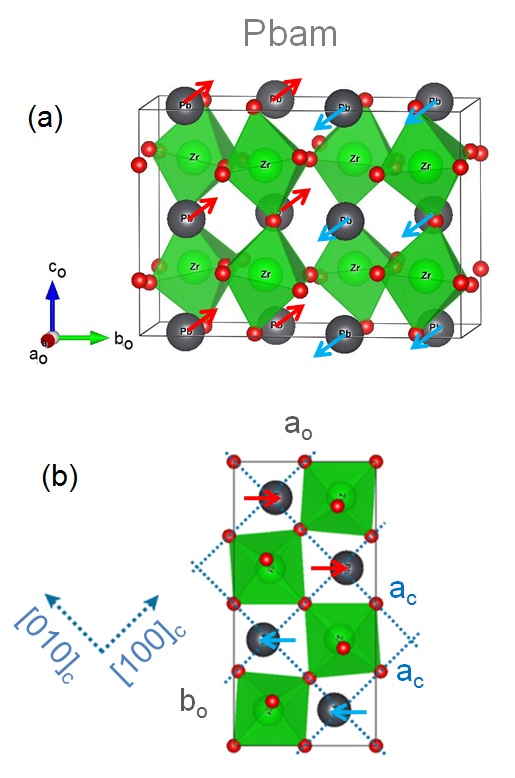} %{Fig6.jpg}
		\caption{\textbf{Structure of PbZrO$_3$ \cite{Izumi2011} }.
			(a) Crystal structure of the $Pbam$ unit cell of PbZrO$_3$. (b) Projection of the $Pbam$ structure onto the $(x,y)$-plane. The arrows indicate the displacements of the Pb cations in the $\Sigma_2$ mode. These Pb displacements along the pseudocubic $[1\bar{1}0]$ direction can be written \cite{Wei2014} as $\textbf{r}_{Pb} \propto (1,-1,0)cos\left(\frac{\pi}{2a_c}(x_c+y_c)+\varphi\right)$.}
		\label{fig:PZO structure}
	\end{figure}

The antiferroelectric (AFE) phase transition in lead zirconate at $T_c = 505~K$ transforms the crystal from the cubic perovskite phase $Pm\bar{3}m$ (Z=1) to the orthorhombic phase $Pbam$ (Z=8), with an eightfold multiplication of the cubic unit cell \cite{Fuji1984,Fuji1997}.\\
	The unit cell (Fig.~\ref{fig:PZO structure}) of the AFE $Pbam$ phase with lattice vectors $\textbf{a}_o, \textbf{b}_o, \textbf{c}_o$ is related to the cubic phase with $\textbf{a}_c=(a,0,0), \textbf{b}_c=(0,a,0), \textbf{c}_c=(0,0,a)$ as $\textbf{a}_o=\textbf{a}_c-\textbf{b}_c$, $\textbf{b}_o=2(\textbf{a}_c+\textbf{b}_c)$ and $\textbf{c}_o=2\textbf{c}_c$, with the origin shifted to $(\frac{a}{2},0,\frac{a}{2})$. 
	This unit cell corresponds to the orthorhombic DSs $1_i$ described below. Other orientational domains are obtained by appropriate rotations. 
	
	The soft mode behaviour \cite{Tagantsev2013,Hlinka2014,Stengel2014} is rather complicated and involves many modes. However, the
	space-group symmetry change can be well understood \cite{Rabe2013,Toledano2019} as a result of the condensation of 2 order parameters. The lead displacements are described by the condensation of a wave with a propagation vector $\textbf{k}_{\Sigma}=\frac{2\pi}{a}(\frac{1}{4},\frac{1}{4},0)$, with the 12-component eigenvector
	\begin{equation}\label{eq:DS}
	\bm{\eta}=\{\eta_1,\eta_2,\eta_3,\eta_4,\eta_5,\eta_6,\eta_7,\eta _8,\eta_9,\eta_{10},\eta_{11},\eta_{12}\}
	\end{equation}
	
	\begin{table}[h]
		\begin{equation}
		\begin{array}{c|cccccccccccc|ccc}
		DSs&\multicolumn{12}{c}{\bm{\eta}}&\multicolumn{3}{|c}{\bm{\phi}}\\
		\hline
		1_i&\eta_1&\eta_2&0 &0 &0 &0 &0 &0 &0 &0 &0 &0 & \multirow{2}{*}{$\phi_1$}&\multirow{2}{*}{$\phi_2$}&\multirow{2}{*}{0} \\
		2_i&0 &0 &\eta_3&\eta_4 &0 &0 &0 &0 &0 &0 &0 &0  \\ \hline
		3_i&0 &0 &0 &0 &\eta_5&\eta_6&0 &0 &0 &0 &0 &0 & \multirow{2}{*}{0}&\multirow{2}{*}{$\phi_2$}&\multirow{2}{*}{$\phi_3$} \\
		4_i&0 &0 &0 &0 &0 &0 &\eta_7&\eta_8&0 &0 &0 &0  \\ \hline
		5_i&0 &0 &0 &0 &0 &0 &0 &0 &\eta_9&\eta_{10}&0 &0 & \multirow{2}{*}{$\phi_1$}&\multirow{2}{*}{0}&\multirow{2}{*}{$\phi_3$} \\
		6_i&0 &0 &0 &0 &0 &0 &0 &0 &0 &0 &\eta_{11}&\eta_{12} \\ \hline
		\end{array}
	\end{equation}
\caption{\label{tab:1} Symmetry allowed OP components $\eta_i$ in the six orientational DSs and corresponding octahedral tilts $\phi_i$.}
	\end{table}
	which transforms according to the irreducible representation \cite{Aroyo2006} $\Sigma_2$. There are 6 orientational domain states (DS) each described by one pair of OP components as shown in TABLE \ref{tab:1}. Since the condensation of $\Sigma_2$ is associated with a fourfold increase in the number of atoms in the unit cell, every orientational DS can exist in 4 different translational DSs with respect to lead displacements \cite{footnote1} numbered by $i$-index in TABLE \ref{tab:1}, $i=1,..,4$. The equilibrium values of the OP pair $(\eta_1,\eta_2)$ are $(\eta,-\eta),\ (-\eta,-\eta),\ (-\eta,\eta),\ (\eta,\eta)$ for DSs $1_1,\ 1_2,\ 1_3,\ 1_4$, respectively. It is analogous for the remaining DSs. 
	
\begin{figure} 
		\centering
		\includegraphics[scale=0.5]{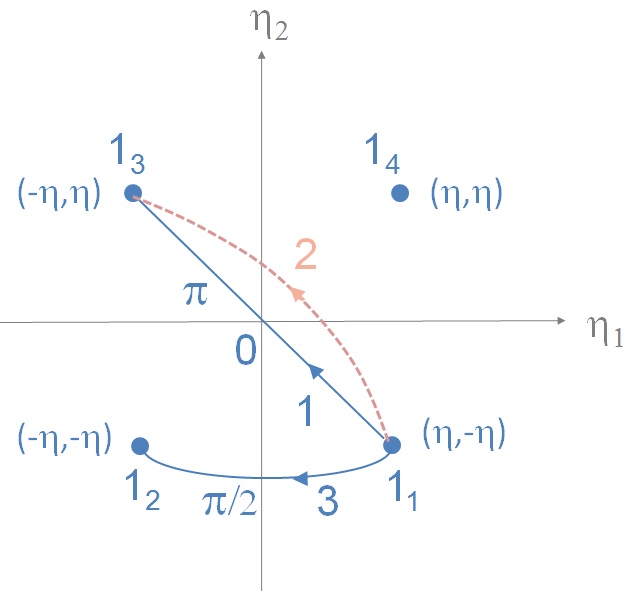} %{Fig7.jpg}
		\caption{\textbf{Translational DSs and APBs in OP-space}.
			Mapping of the translational domain states and APBs within a single orientational DS of PbZrO$_3$ in the subspace $(\eta_1,\eta_2)$ of the OP-space $\bm \eta$. Circles represent the 4 different domain states $1_1, 1_2, 1_3, 1_4$ and lines describe some possible pathways of the corresponding APB in the OP-space. E.g. line 1 represents an Ising wall between $1_1$ and $1_3$ and line 2 is an 'improper' Ne\'el wall of the corresponding domain pair.}
		\label{fig:PZO domain states}
	\end{figure}

	The translational DSs within a given orientational state can be transformed from one to another by shifting the lattice by lost translations, e.g. DSs $1_i$ are related by $(a,0,0), (2a,0,0)$ or $(3a,0,0)$, which after application of the matrix element $(1 \vert t_1,t_2,t_3)$, $(t_1=na, t_2=ma, t_3=la)$ of the irrep $\Sigma_2$ leads to the identification 
	
	\begin{eqnarray}
		\label{shift of DSs in lead displacements}
		1_1 - (\mathbf{a_c}, \pi/2) \rightarrow 1_2  \nonumber\\ 
		1_1 - (2\mathbf{a_c}, \pi) \rightarrow 1_3  \\ 
		1_1 - (3\mathbf{a_c}, 3\pi/2) \rightarrow 1_4  \nonumber
	\end{eqnarray}
	
	where it is shown that the corresponding lead displacement modes in different translatioinal DSs are phase-shifted by $\Delta \varphi = \pi/2, \pi$ and $3\pi/2$, respectively (Fig.~\ref{fig:PZO domain states}). 
	The orientational DSs are related by rotations, DSs '1' and '2' by $\pi/2$ rotation about $\mathbf{c}_c$, '3' and '4' about $\mathbf{a}_c$, '5' and '6' about $\mathbf{b}_c$. The AFE lead displacements are along cubic (1,-1,0) in $1_i$ and (1,1,0) in $2_i$ DSs, (1,0,-1) in $3_i$ and (1,0,1) in $4_i$ DSs, (0,1,-1) in $5_i$ and (0,1,1) in $6_i$ DSs.
	
	 A condensation of the $\Sigma_2$-mode alone, would lead \cite{Toledano2019} to a $Pbam(4V)$-phase, with a change of the unit cell volume by a factor of 4. In order to get the  $Pbam(8V)$-phase, one has to take into account the antiphase rotations of the oxygen octahedra, which are described by the condensation of a wave with $\textbf{k}_{R}=\frac{2\pi}{a}(\frac{1}{2},\frac{1}{2},\frac{1}{2})$, with order parameter $\bm{\phi}=(\phi_1,\phi_2,\phi_3)$ transforming under the irrep $R_4^+$ and yielding a doubling of the unit cell in $c$-direction. $\phi_1,\phi_2,\phi_3$ represent octahedra rotations about $x,y,z$, respectively. The equilibrium OP for PbZrO$_3$ has always one zero component, and it has 2 possible values, e.~g. for DSs $1_i$ and $2_i$, $(\phi_1,\phi_2,0)=(\phi,-\phi,0)$ or $(-\phi,\phi,0)$, see TABLE~\ref{tab:1}.\\
	The symmetry breaking $Pm\bar{3}m (V) \rightarrow Pbam (8V)$ leads to $48/8 \times 8 = 48$ DSs. Further the symmetry of the $\pi/2$ and $\pi$ APBs is studied in order to find out the APB's polar properties and compare them with recent experimental results and simulations \cite{Wei2014,Wei2015}. We use the approach based on the layer group analysis and Landau theory as it was suggested in previous papers. 
	
	\begin{figure} 
		\centering
		\includegraphics[scale=0.5]{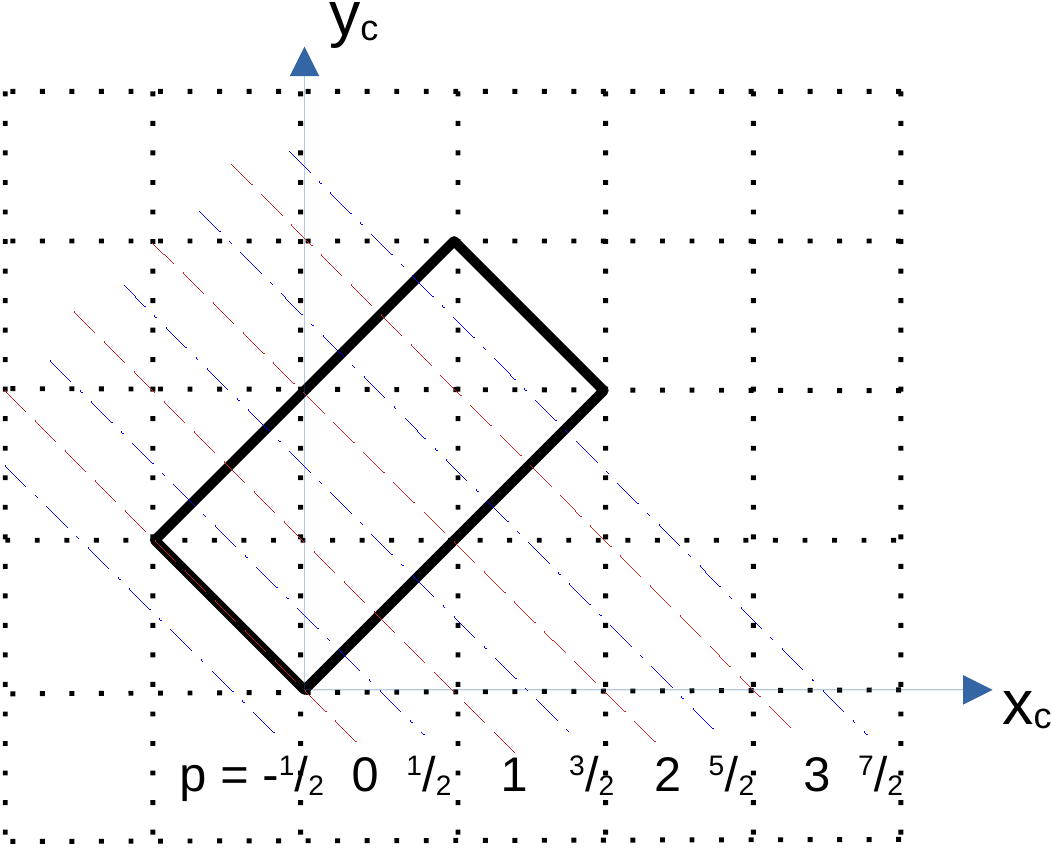} %{wall_position-cropped.pdf}
		\caption{\textbf{Unit cell of the $1_1$ DS and \textit{potentially} significant positions of the APB}.
		The position $p$ is determined by the intersection of the APB plane (color lines) with the $x-$axis. The set of blue DWs (if any) possess the same point symmetry, and the same applies to the set of red DWs (examples are in Sections \ref*{sec:11|12} and \ref*{sec:11|13}). The cubic coordinates are used. }
		\label{fig:APB_positions}
	\end{figure}

\section{Symmetries of objects related to DW\lowercase{s}}\label{sec:symm-define}
The DW's symmetry is closely related to other mathematical objects such as domain states, domain pairs, domain twins, etc., and for further calculations it is convenient to introduce the following sequence of objects built from domain states $A$ and $B$ with lowering symmetry: 
Symmetry of unordered domain pair (DP) $\supset$ symmetry of oriented DP $\supset$ union of symmetries of DWs at all positions $\supset$ symmetry of a single DW at position $p$. It can be formally written as:
\begin{eqnarray}\label{eq:basics1}
	\lefteqn{(A,B)+(B,A)\supset (A,B)(\mathbf{n})+(B,A)(\mathbf{-n})} \nonumber \\ 
	&\supset  \sum_{p'}\left((A,B)(\mathbf{n})+(B,A)(\mathbf{-n})\right)(\mathbf{p'}) 
	\supset \left((A,B)(\mathbf{n})+(B,A)(\mathbf{-n})\right)(\mathbf{p})
\end{eqnarray}
and in short notation:
\begin{equation}\label{eq:basics2}
	S_0\equiv \{A,B\}\supset S_1\equiv(A|\mathbf{n}|B)\supset  S_2\equiv\sum_{p'}(A|\mathbf{n,p'}|B)\supset S_3\equiv(A|\mathbf{n,p}|B)
\end{equation}	
The symmetries $S_1,\ S_2,\ S_3$ are used in the next sections. Note that the above description is related to the DW of 2 DSs $A$ and $B$, separated by a plane with the normal vector $\mathbf{n}$ and position  $\mathbf{p}$. Since such a wall contains only 2 domains, we refer to it as a simple DW. \\  
In more complex DWs a precursor (nucleus) of an additional DS, say $C$, may occur at the DW center \cite{Salje1991,Bullbich1989}. The symmetry operations of $C$ yield further restrictions on symmetry and it is encountered in expression (\ref{eq:basics1}) as:
\begin{eqnarray}\label{eq:basics10}
	\lefteqn{((A,B)+(B,A))(C)\supset ((A,B)(\mathbf{n})+(B,A)(\mathbf{-n}))(C)}
	 \nonumber \\ &\supset\left(\sum_{p'}\left((A,B)(\mathbf{n})+(B,A)(\mathbf{-n})\right)(\mathbf{p'})\right)(C)\supset  \left((A,B)(\mathbf{n})+(B,A)(\mathbf{-n})\right)(C)(\mathbf{p})
\end{eqnarray}
Note that the last expression in (\ref{eq:basics10}) in principle describes the DW between $A$ and $B$ separated by the plane with normal $\mathbf{n}$ and position $\mathbf{p}$, and with $C$ at the center. Its short notation can be $(A|C,\mathbf{n,p}|B)$ (compare with (\ref{eq:basics2})). It should be stressed that we always consider a single DW with one plane $\mathbf{n}$ and one position $\mathbf{p}$. For comparison note, that one can also study a structure of 2 separated DWs: $(A|\mathbf{n_1,p_1}|C)+(C|\mathbf{n_2,p_2}|B)$. Here $C$ denotes a fully developed DS, while $C$ above means a 'nucleus' of $C$ at the center. In next Sections it will be also shown that the symmetry $S_2$ is reflected in the free energy gradient invariants, while the microscopic positions of DWs inside the unit cell are 'invisible' by the Landau-Ginzburg approach. 

In order to avoid misunderstanding we further use the following naming of the APBs. The APB $(A|B)$ composed of 2 translational DSs is mentioned as the Ising-like. The APB $(A|C|B)$ with nucleus (precursor) of the translational  domain state $C$ with respect to $A$ and $B$ is called an improper N\'eel-like (see the path '2' in Fig.~\ref{fig:PZO domain states}).  If $C$ represents a different orientational DS, then $(A|C|B)$ is called a (proper) N\'eel structure. 

\section{On the polarization in $(1_1|1_2)$ APB}\label{sec:11|12}
\begin{figure}[h!]
	\centering
\begin{subfigure}[b]{0.4\textwidth}
	\includegraphics[scale=0.4]{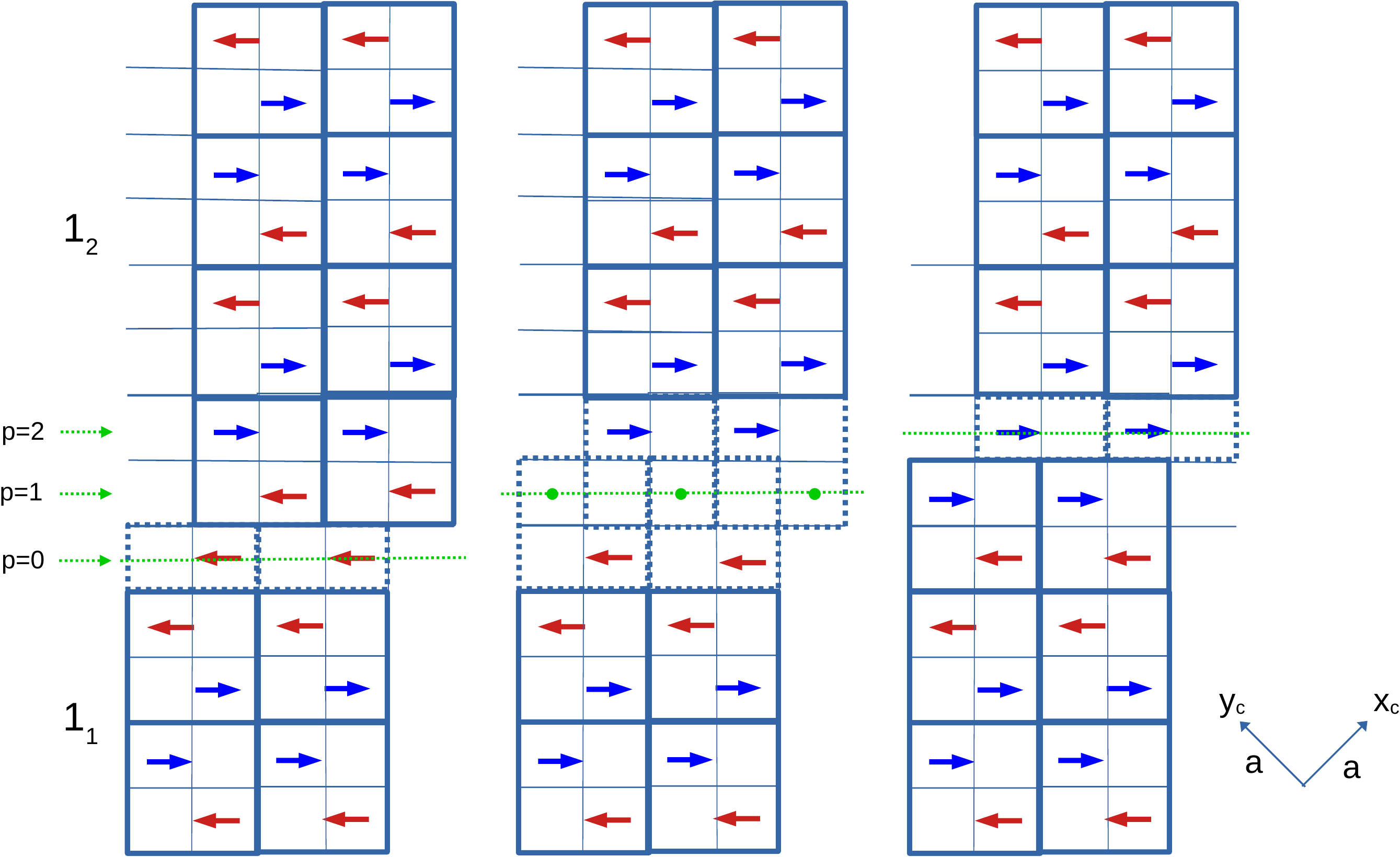} %{pzo_domains_1112.pdf}
	\caption{}
\end{subfigure}
\hfill
\begin{subfigure}[b]{0.33\textwidth}
	\includegraphics[scale=0.25]{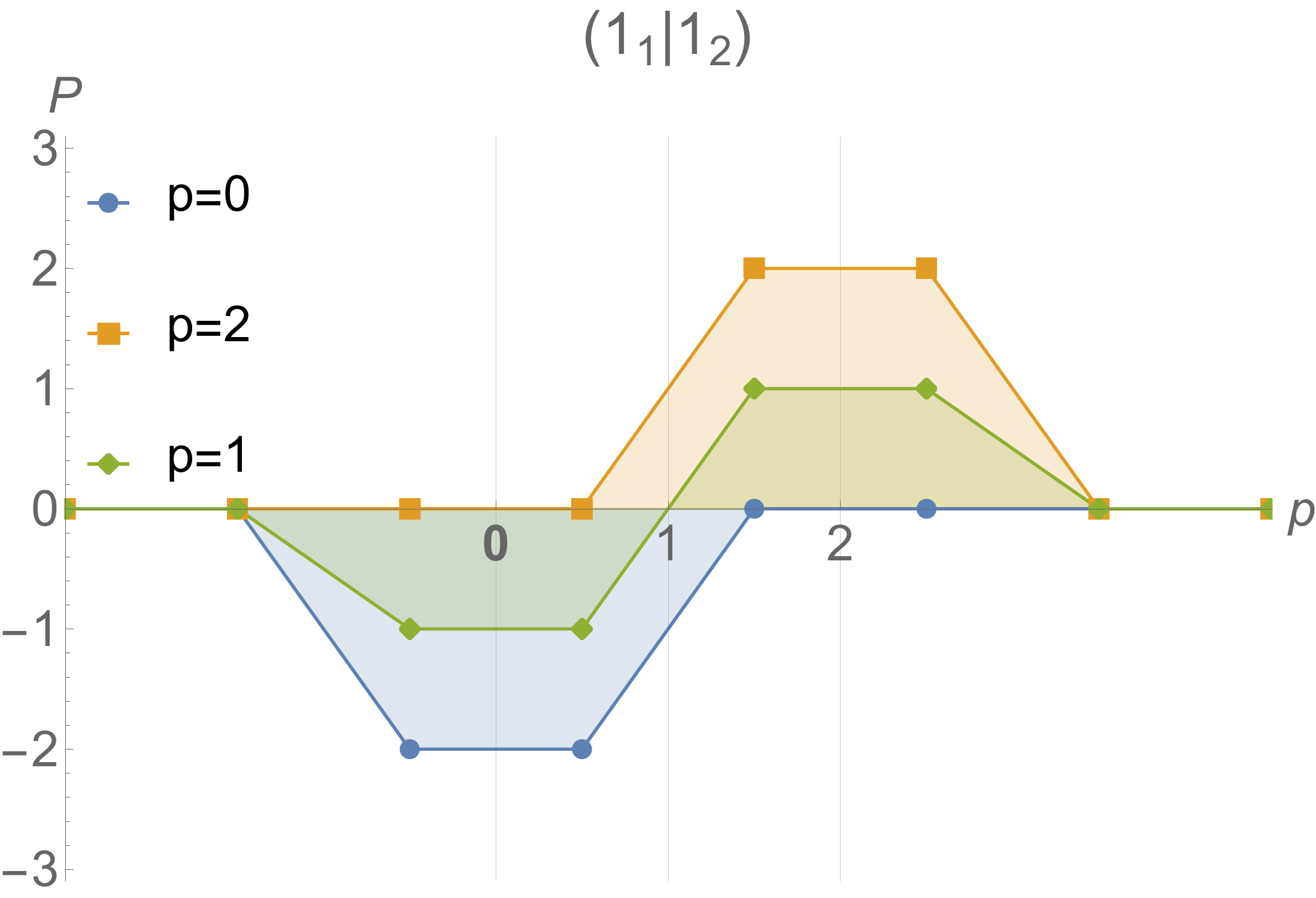} %{sliding_profil3_big.pdf}
	\caption{}
\end{subfigure}	 
	\caption{\textbf{ $\pi/2$--APB $(1_1|1_2)$ (Ising-like type) at different positions. } (a) The APBs at $p=0$ and $p=2$ are symmetry related and have opposite polarizations. The APB at $p=1$ has higher symmetry with an antisymmetric polarization profile.  
	(b) The polarization profiles of the APBs were obtained by 'sliding unit cell' method. The macroscopic polarization is an average over all positions and yields 0.}
	\label{fig:1112}
\end{figure}

Let us consider the $\pi/2-$type APB between the DSs $1_1$ and $1_2$ (Fig.~\ref{fig:PZO domain states}), which were defined in the previous section. $1_1=(\eta ,-\eta ,0,\dots,0)$ and 
$1_2=(-\eta ,-\eta ,0,\dots0)$, both 
with octahedra rotations $(\phi ,-\phi,0)$, see TABLE~\ref{tab:1}. The normal of the DW plane $\mathbf{n}$ can be taken arbitrary, here we consider $\mathbf{n}=(1,1,0)$ since  it was recently discussed by several authors. The microscopic position of the APB inside the unit cell is denoted as $\mathbf{p}$ and for the moment it is left arbitrary. All possible positions and corresponding symmetries will be generated during the calculation, Fig~\ref{fig:APB_positions}. The APB between DSs $1_1$ and $1_2$, with the normal $\mathbf{n}$ and position $\mathbf{p}$ is denoted as $(1_1|\mathbf{n,p}|1_2)$. For further analysis we consider the symmetries of the following 3 objects, (see (\ref{eq:basics2})):  \\
\begin{enumerate}[label=(\roman*)]
	\item The symmetry $S_1$ of the object $(1_1|\mathbf{n}|1_2)$ (note that $(1_1|\mathbf{n}|1_2)\equiv (1_2|\mathbf{-n}|1_1)$) consists of all operations, which either leave the DSs and the normal invariant, or interchange DSs and at the same time  reverse the normal vector. All operation from $S_1$ keep the orientation of the DW plane fixed, but some of them can move the plane along the normal $\mathbf{n}$. 
	\item Symmetry $S_2$ of object $(1_1|\mathbf{n,p}_{arb}|1_2)$, where the arbitrary position $\mathbf{p}_{arb}$ of the APB is considered (all possible positions are taken into account). $S_2$ is obtained from $S_1$, by choosing only operations, which do not move the DW plane along $\mathbf{n}$. The position of the DW is defined as the point, where the plane intersects the $x$-axis, $\mathbf{p}=(p,0,0)$, see Fig.~\ref{fig:APB_positions}. The point $(x,y,z)$ lies in the plane positioned at $\mathbf{p}=(p,0,0)$, when $\mathbf{n}\cdot (x,y,z)=\mathbf{n}\cdot \mathbf{p}$. In our case it yields $x+y=p$. The transformed point $(x',y',z')$ lies in the same plane if $\mathbf{n}\cdot (x',y',z')=\mathbf{n}\cdot \mathbf{p}$, i.~e., $x'+y'=p$. We can apply these equations to the elements of $S_1$ (see (iii) below) and obtain the positions $p$. E.~g., for the first operation the equation of the plane reads $(-x+1)+(-y+1)=p$, and since $x+y=p$, it yields the position $p=1$. The positions of planes invariant with respect to individual operations are listed in (\ref{eq:sym1}) just behind $S_1$. $p=p$ means that any position is possible, while $p='-'$ indicates no solution, i.~.e., the last 2 operations shift the plane along $\mathbf{n}$. Now it is clear that $S_2$ corresponds to the union of symmetry operations of APBs at all positions. In this way we automatically obtained all possible positions of the APBs, 2 special positions $p=0,1$ (high symmetric) and a general position $p$ (low symmetric, neither 0 nor 1), see below. The 'average' symmetry $S_2$ determines, whether a macroscopic polarization can exist or not. Since its point group contains the inversion element, the macroscopic polarization does not exist, and therefore there are no Landau free energy expansion terms, which could generate it (see next section). 
	\item The microscopic symmetries $S_3$ (\textit{the layer groups}) of the APB $(1_1|\mathbf{n,p}|1_2)$ at particular microscopic positions $p$ are obtained from $S_2$ by extracting the operations with particular values of $p$.
\end{enumerate}
\begin{equation}\label{eq:sym1}
S_1=
\left(
\begin{array}{ccc}
	\bar{x}+1 & \bar{y}+1 & \bar{z} \\
	\bar{x}+1 & \bar{y}+1 & z+1 \\
	x & y & \bar{z}+1 \\
	x & y & z \\
	\bar{y} & \bar{x} & \bar{z} \\
	\bar{y} & \bar{x} & z+1 \\
	y+1 & x+1 & \bar{z}+1 \\
	y+1 & x+1 & z \\
\end{array}
\right)
p=
\left(
\begin{array}{ccc}
	1 \\
	1 \\
	p \\
	p \\
	0 \\
	0 \\
	- \\
	- \\
\end{array}
\right)
\quad
S_2 = 
\left(
\begin{array}{ccc}
	\bar{x}+1 & \bar{y}+1 & \bar{z} \\
	\bar{x}+1 & \bar{y}+1 & z+1 \\
	x & y & \bar{z}+1 \\
	x & y & z \\
	\bar{y} & \bar{x} & \bar{z} \\
	\bar{y} & \bar{x} & z+1 \\
\end{array}
\right)
p=
\left(
\begin{array}{ccc}
	1 \\
	1 \\
	p \\
	p \\
	0 \\
	0 \\
\end{array}
\right)
\end{equation}
\begin{equation}\label{eq:sym2}
S_3(p=0) = 
\left(
\begin{array}{ccc}
	\bar{y} & \bar{x} & \bar{z} \\
	\bar{y} & \bar{x} & z+1 \\
	x & y & \bar{z}+1 \\
	x & y & z \\
\end{array}
\right)
\quad
S_3(p=1) = 
\left(
\begin{array}{ccc}
	\bar{x}+1 & \bar{y}+1 & \bar{z} \\
	\bar{x}+1 & \bar{y}+1 & z+1 \\
	x & y & \bar{z}+1 \\
	x & y & z \\
\end{array}
\right)
\quad
S_3(p=gen) = 
\left(
\begin{array}{ccc}
 	x & y & \bar{z}+1 \\
	x & y & z \\
\end{array}
\right)
\end{equation}
$S_3(p=0)$ allows $\mathbf{P}=(P,-P,0)$; $S_3(p=1)$ is nonpolar; $S_2$ is average symmetry (we can call it 'macroscopic') and it is {\bfseries nonpolar} in agreement with invariants, as will be shown below.\\
Allowed translations corresponding to $S_1,S_2,S_3$ are:\\
$\mathbf{T_1}=(k(0, 0, 2), l(1, -1, 0), m(2, 2, 0)))$\\
$\mathbf{T_2}=\mathbf{T_3}=(k(0, 0, 2), l(1, -1, 0)))$\\
Let us note that in general there are symmetrically equivalent APBs at positions $p=\dots, -3,-1,+1,+3,\dots$, and another set of APBs at $p=\dots, -2,0,+2,\dots$. The symmetry operation $\bar{x}+1,\bar{y}+1,z+1$ of APB($p=1$) transforms the APB($p=0$) with $+\mathbf{P}$ to the APB($p=2$) with reversed polarization $-\mathbf{P}$. A similar effect have the last 2 operations in the matrix $S_1$. 
The graphical representation of the Ising-like APB $(1_1|1_2)$ at the possible positions discussed before is shown in Fig.~\ref{fig:1112}. The polarization at the center is reversed when shifting the APB from $p=0$ to $p=2$. The polarization profiles at these positions have the form of a single peak, while the profile at $p=1$ has an antisymmetric shape. The polarization averaged over all positions is zero as predicted by $S_2$. Let us mention in advance that the symmetry $S_2$ also indicates that the macroscopic polarization profile obtained from the free energy description is exactly zero, see the curve '1' in Fig.~\ref{fig:APB_profiles_P}.

In order to get a macroscopic polarization an additional symmetry lowering of $(1_1|\mathbf{n}|1_2)$ is needed. It can be achieved in the APB of the N\'eel-type with a nucleus of another DS at the center, which can occur due to a local phase transition in the DW \cite{Bullbich1989,Kvasov2016,Salje1991}. We consider $(1_1|2_1,\mathbf{n,p}|1_2)$, i.~e., there is in addition the nucleus $2_1$ (orientational DS) at the APB center, see Section \ref{sec:symm-define} and comments therein. 
The procedure outlined above yields:    
\begin{equation}\label{eq:sym3}
S_1=S_2=S_3(p=0) = 
\left(
\begin{array}{ccc}
	x & y & \bar{z}+1 \\
	x & y & z \\
	\bar{y}-1 & \bar{x}+1 & \bar{z} \\
	\bar{y}-1 & \bar{x}+1 & z+1 \\
\end{array}
\right)
p=
\left(
\begin{array}{ccc}
	p \\
	p \\
	0 \\
	0 \\
\end{array}
\right)
\end{equation}
and $S_3(p=gen)$ is the same as in (\ref{eq:sym2}). 
In this case the APB is (macroscopically) {\bfseries polar}, $\mathbf{P}=(P,-P,0)$, and note that there is only one set of APB's positions with high symmetry, $p = ..., -2, 0, 2, ...$. The APBs at $p=0$ and $p=2$ have the same symmetry allowing polarization $(P,-P,0)$, but they are not related by any operation and therefore their polarization vectors cannot cancel.\\
Allowed translations corresponding to $S_1,S_2,S_3$ are:\\
$\mathbf{T_1}=(k(0, 0, 2), l(2, -2, 0), m(2, 2, 0)))$\\
$\mathbf{T_2}=\mathbf{T_3}=(k(0, 0, 2), l(2, -2, 0)))$\\
To be more precise and avoid misunderstanding concerning equations $(\ref{eq:sym3})$ let us note that the full symmetries are related as:\\
$(S_1\oplus\mathbf{T_1})\supset (S_2\oplus\mathbf{T_2})=(S_3(p=0)\oplus\mathbf{T_3}) \supset (S_3(p=gen)\oplus\mathbf{T_3})$.\\

\begin{figure}[h!] 
	\centering
\begin{subfigure}[b]{.42\textwidth}
	\includegraphics[scale=0.4]{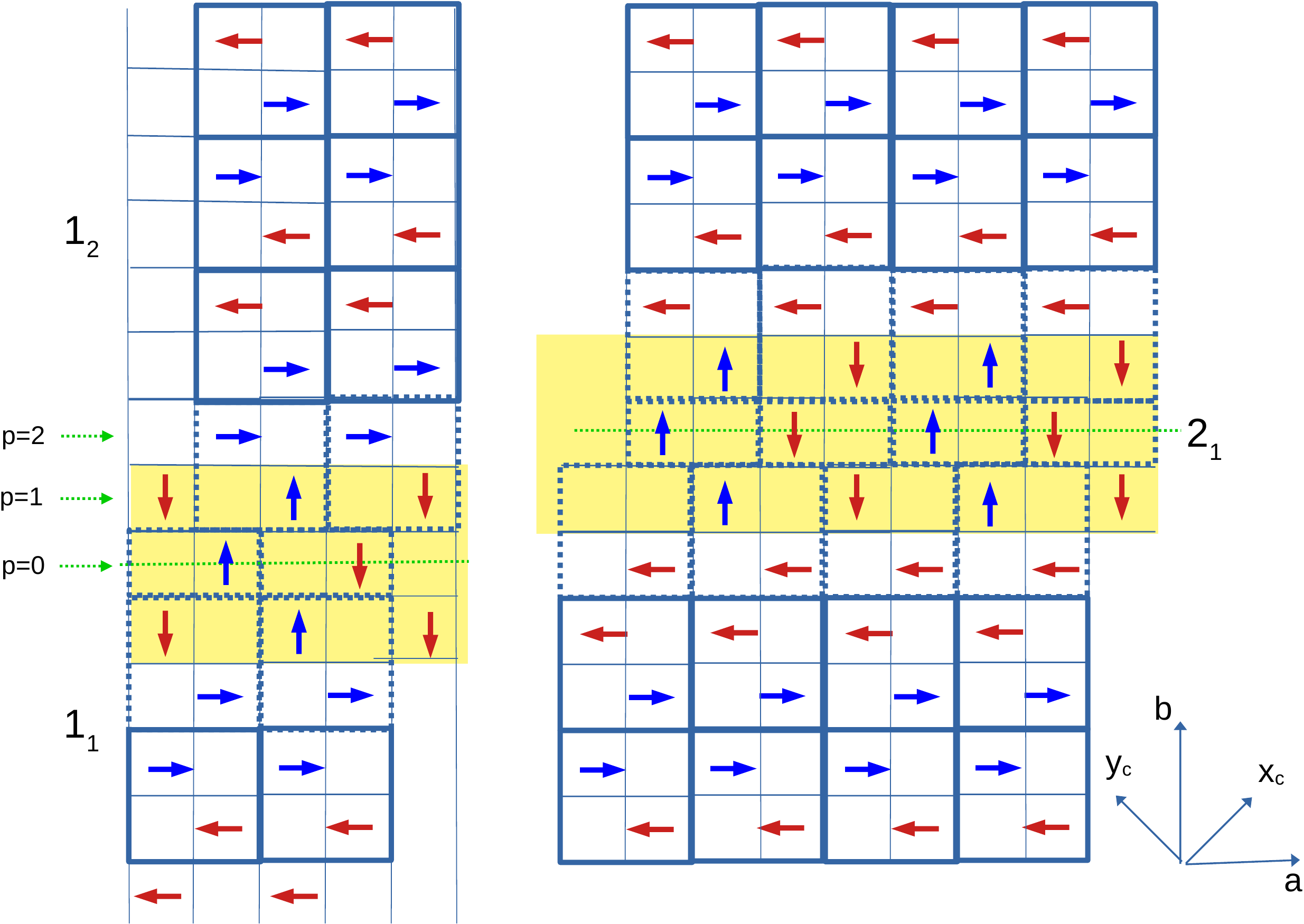} %{pzo_domains_112112.pdf}
\caption{}
\end{subfigure}
 \hfill
 %\hspace{25mm}
\begin{subfigure}[b]{.4\textwidth}
	\includegraphics[scale=0.4]{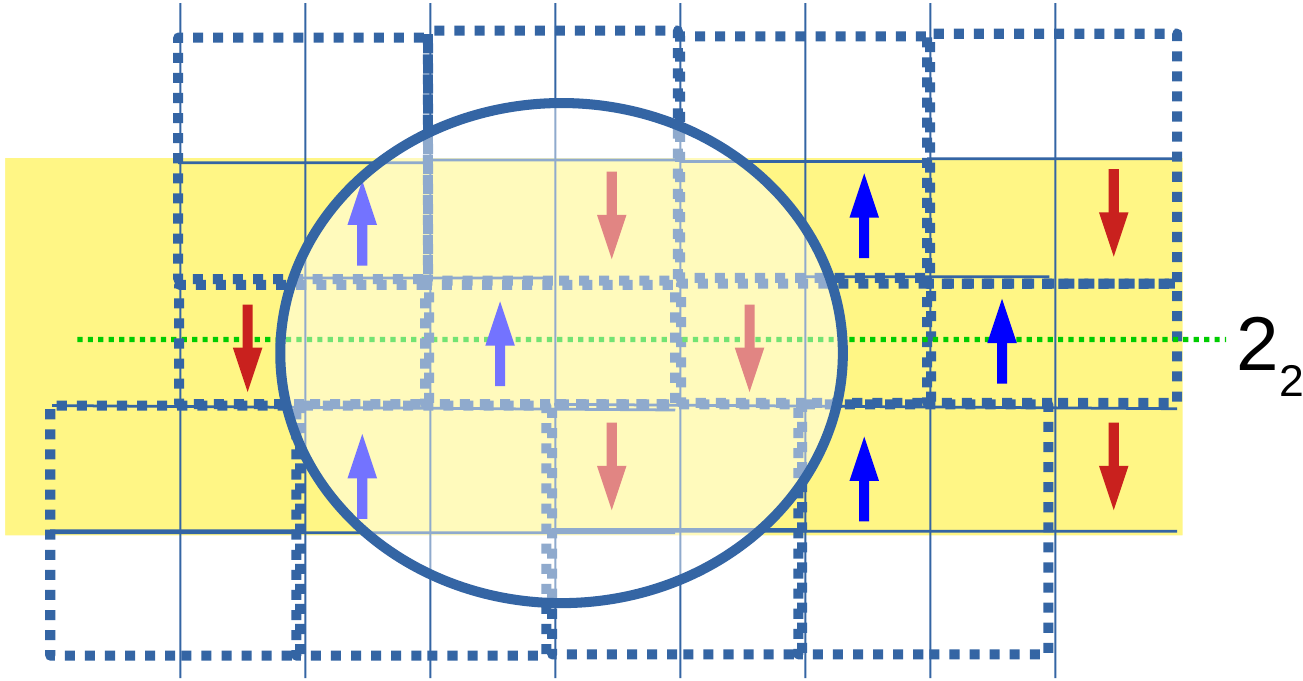}%{pzo_domains_112212_core2.pdf}
	\caption{}
	\vspace{5mm}
	\includegraphics[scale=0.25]{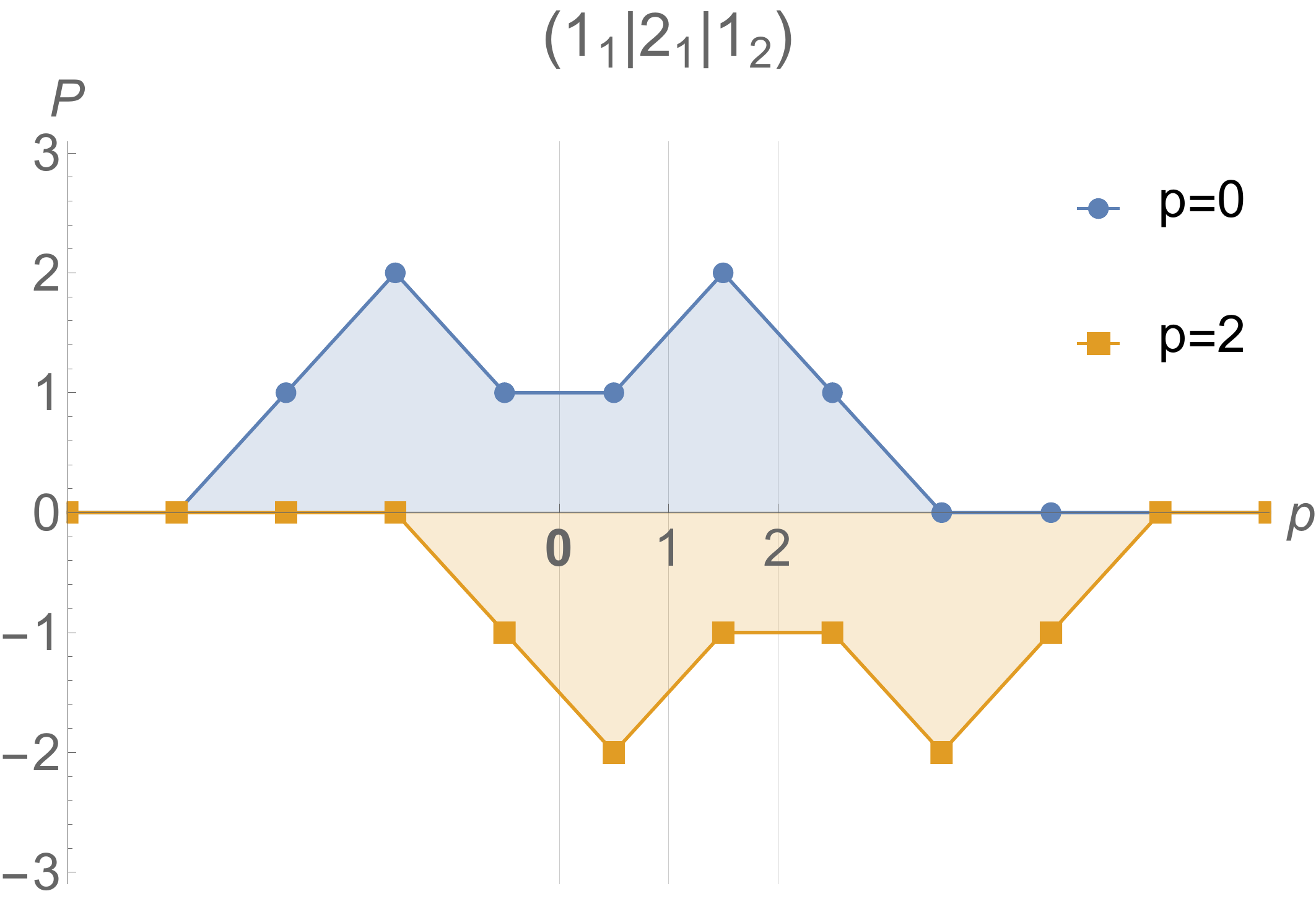}%{sliding_profil2_big.pdf}
	\caption{}
\end{subfigure}
	\caption{\textbf{$\pi/2$--APB with a nucleus: $(1_1|2_1|1_2)$ (N\'eel-like type).} (a): There are only two high-symmetric positions $p=0,\ 2$ (see dotted green lines) as compare with Fig.~\ref{fig:1112}, and they are not related by any symmetry operation. The nucleus '$2_1$' shows the left oriented wedge pattern. (b) The nucleus '$2_2$' with the right oriented wedge pattern. The switching from $2_1$ to $2_2$ in (a) is accomplished by the reversal of arrows in the central row. (c) The sliding unit cell calculation of the polarization profiles of $(1_1|2_1|1_2)$. The double peak reflects the presence of two interfaces, $1_1|2_1$ and $2_1|1_2$ at the center of the wall. The polarizations at $p=0$ and $p=2$ are still opposite and seemingly equal as in Fig.~\ref{fig:1112}, but in reality they differ, since symmetry $S_2$ dictates a nonzero average of in-plane polarization $P$. }
	\label{fig:112112}
\end{figure}

The N\'eel-like APB $(1_1|2_1|1_2)$ with the nucleus $2_1$ is illustrated in Fig.~\ref{fig:112112}. Note that the polarizations at positions $p=0$ and $p=2$ have again opposite signs, but the mean value is nonzero. It means a non-zero macroscopic polarization, qualitatively shown in curve '3' of Fig.~\ref{fig:APB_profiles_P}).  
\begin{figure}[h!] 
	\centering
	\includegraphics[scale=0.5]{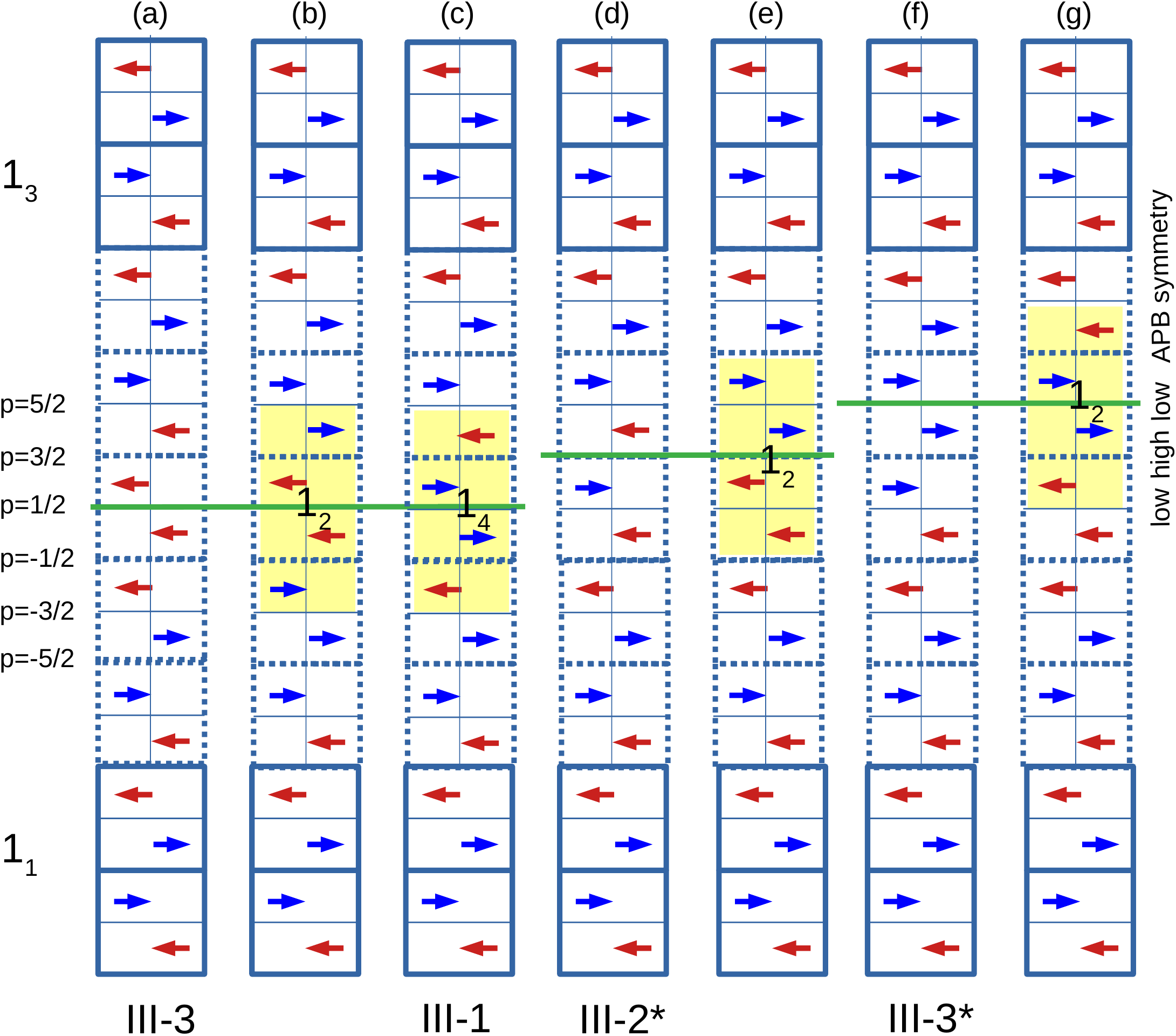}%{pzo_domains_1113__2.pdf}
	\includegraphics[scale=0.2]{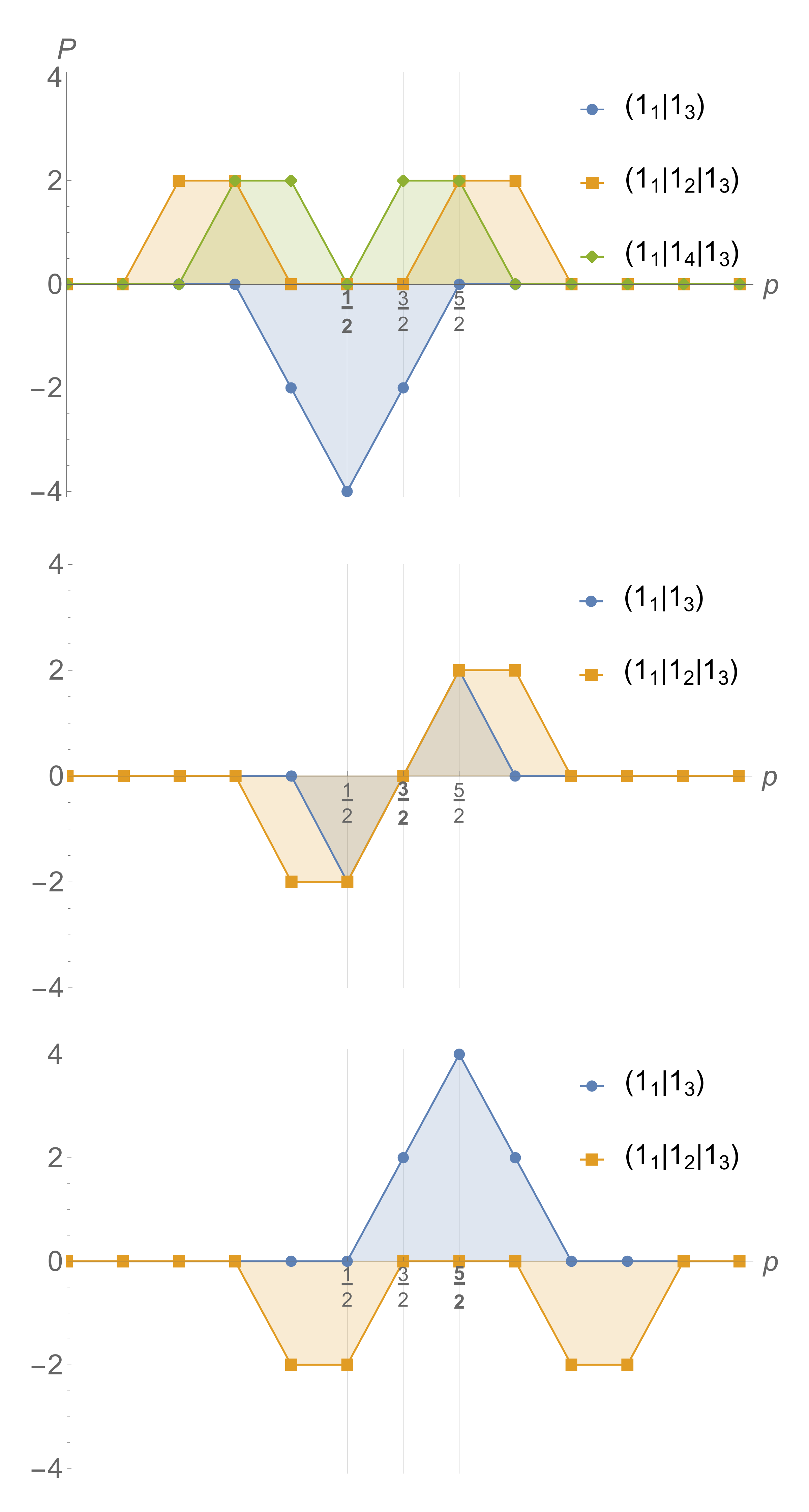}%{sliding_profil_big.pdf}
	\caption{\textbf{Selected $\pi$--APBs.} Left: The green lines show APBs' positions. The APBs discussed in Refs.~\cite{Wei2014,Wei2015} are indicated by the symbols at the bottom. The symbol '*' denotes a symmetrically equivalent APBs at different positions than in Ref.~\cite{Wei2014,Wei2015} , e.~g., III-2 is at $p=-1/2$, while III-2* is at $p=3/2$. Right: The polarization profiles of the APBs were obtained by 'sliding unit cell' method. The double peak is caused by a nucleus at the APB's center.     }
	\label{fig:1113}
\end{figure}
\section{Symmetry of $(1_1|1_3)$ APB}\label{sec:11|13}
Here, the procedure outlined in the previous section is used to study the symmetry and polarity of the APBs between $1_1$ and $1_3$ DSs with Ising-like and N\'eel-like structures (Fig.~\ref{fig:PZO domain states}). The symmetries related to the Ising-like $(1_1|1_3)$ APB read:
\begin{equation}\label{eq:sym13_1}
	S_1=
	\left(
	\begin{array}{ccc}
		\bar{x}+1 & \bar{y}+2 & \bar{z}+1 \\
		\bar{x}+1 & \bar{y}+2 & z \\
		x & y & \bar{z}+1 \\
		x & y & z \\
		\bar{y}+1 & \bar{x} & \bar{z}+1 \\
		\bar{y}+1 & \bar{x} & z \\
		y+1 & x+1 & \bar{z}+1 \\
		y+1 & x+1 & z \\
	\end{array}
	\right)
	p=
	\left(
	\begin{array}{ccc}
		3/2 \\
		3/2 \\
		p \\
		p \\
		1/2 \\
		1/2 \\
		- \\
		- \\
	\end{array}
	\right)
	%\quad
	S_2 = 
	\left(
	\begin{array}{ccc}
		\bar{x}+1 & \bar{y}+2 & \bar{z}+1 \\
		\bar{x}+1 & \bar{y}+2 & z \\
		x & y & \bar{z}+1 \\
		x & y & z \\
		\bar{y}+1 & \bar{x} & \bar{z}+1 \\
		\bar{y}+1 & \bar{x} & z \\
	\end{array}
	\right)
	p=
	\left(
	\begin{array}{ccc}
		3/2 \\
		3/2 \\
		p \\
		p \\
		1/2 \\
		1/2 \\
	\end{array}
	\right)
\end{equation}
\begin{equation}\label{eq:sym13_2}
	S_3(p=1/2) = 
	\left(
	\begin{array}{ccc}
		\bar{y}+1 & \bar{x} & \bar{z}+1 \\
		\bar{y}+1 & \bar{x} & z \\
		x & y & \bar{z}+1 \\
		x & y & z \\
	\end{array}
	\right)
	%\quad
	S_3(p=3/2) = 
	\left(
	\begin{array}{ccc}
		\bar{x}+1 & \bar{y}+2 & \bar{z}+1 \\
		\bar{x}+1 & \bar{y}+2 & z \\
		x & y & \bar{z}+1 \\
		x & y & z \\
	\end{array}
	\right)
\end{equation}
	%\quad
\begin{equation*}
	S_3(p=gen) = 
	\left(
	\begin{array}{ccc}
		x & y & \bar{z}+1 \\
		x & y & z \\
	\end{array}
	\right) \nonumber
\end{equation*}
$\mathbf{T_1}=(k(0, 0, 2), l(1, -1, 0), m(2, 2, 0)))$\\
$\mathbf{T_2}=\mathbf{T_3}=(k(0, 0, 2), l(1, -1, 0)))$\\
The APBs at positions $p = \dots, -\frac{3}{2}, \frac{1}{2}, \frac{5}{2}, \dots$ have the same symmetry and allow the polarization along $(1,-1,0)$, but with alternating sign. E.~g., if the APB($p=\frac{1}{2}$) has polarization $(P,-P,0)$, then the APB($p=\frac{5}{2}$), which is obtained from APB($p=\frac{1}{2}$) using either of the 2 last operations in $S_1$, e.~g. $(y+1,x+1,z)$ (or using  ($\bar{x}+1,\bar{y}+1,z$) from $S_2$) possesses polarization $(-P,P,0)$. All the APBs at $p = \dots, -\frac{1}{2}, \frac{3}{2}, \frac{7}{2}, \dots$ are nonpolar (antisymmetric profiles). The corresponding Ising-like APBs at the 3 significant positions are depicted as (a), (d), and (f) in Fig.~\ref{fig:1113}, together with their microscopic polarization profiles. The macroscopic polarization profile determined by the $S_2$ symmetry is identically zero, as shown by the curve '1' in Fig.~\ref{fig:APB_profiles_P}.  

\subsection{Symmetry of $(1_1|1_2|1_3)$ = Symmetry of $(1_1|1_3)$}
Interestingly enough, it turns out that the symmetry properties of the improper N\'eel-type APB $(1_1|1_2|1_3)$ are identical with the Ising-like $(1_1|1_3)$. Its structures at 3 symmetric positions are illustrated as pictures (b), (e), (g) in Fig.~\ref{fig:1113}. Note that the polarization profiles reveal a double-peak due to the 2 interphases $1_1|1_2$ and $1_2|1_3$ nearby the APB's center. The macroscopic polarization profile is, as in the previous case $(1_1|1_3)$, identically zero.  
\subsection{Symmetry of $(1_1|2_2|1_3)$}
The appearance of the orientational/translational nucleus $2_2$ at the center of $(1_1|1_3)$, with the Pb-displacements along $(1,1,0)$ results in the symmetry lowering, the only  high-symmetry positions $p=\dots,-\frac{1}{2},\frac{3}{2},\frac{7}{2},\dots$ correspond to the non-polar APBs, compare with (\ref{eq:sym13_1}) and (\ref{eq:sym13_2}).
\begin{equation}\label{eq:sym123_1}
	S_1=S_2=S_3(p=3/2) = 
\left(
\begin{array}{ccc}
	\bar{x}+1 & \bar{y}+2 & \bar{z}+1 \\
	\bar{x}+1 & \bar{y}+2 & z \\
	x & y & \bar{z}+1 \\
	x & y & z \\
\end{array}
\right)
\quad
	p=
	\left(
	\begin{array}{ccc}
		3/2 \\
		3/2 \\
		p \\
		p \\
	\end{array}
	\right)
\end{equation}
\begin{equation}\label{eq:sym123_2}
	S_3(p=gen) = 
	\left(
	\begin{array}{ccc}
		x & y & \bar{z}+1 \\
		x & y & z \\
	\end{array}
	\right)
\end{equation}
$\mathbf{T_1}=(k(0, 0, 2), l(2, -2, 0), m(2, 2, 0)))$\\
$\mathbf{T_2}=\mathbf{T_3}=(k(0, 0, 2), l(2, -2, 0)))$\\
Note that there is only one high-symmetry position at $p=3/2$. We do not plot the structure of this APB explicitly, but one can find that the microscopic 'sliding cell' profile is antisymmetric (it is qualitatively similar with the profile of $(1_1|1_3)$ at $p=3/2$ in Fig.~\ref{fig:1113}). It is interesting that in this case the symmetry $S_2$ allows the macroscopic polarization profile, which is antisymmetric, see the curve '2' in Fig.~\ref{fig:APB_profiles_P}, Sec.~\ref{sec:freenergy}. 

\subsection{Symmetry of $(1_1|3_2|1_3)$}
Note that the Pb-displacements in all DSs $1_1,1_2,1_3,2_2$ considered so far in the APBs lie in the $xy$ plane. Unlike the previous walls, the nucleus $3_2$ has the Pb-displacements lying in the $yz$  plane, inclined $\pi/4$ rad from $xy$ plane. The symmetries are
\begin{equation}\label{eq:sym133_1}
	S_1=S_2=S_3(p=3/2) = 
	\left(
	\begin{array}{ccc}
		\bar{x}+2 & \bar{y}+1 & \bar{z}+1 \\
		x & y & z \\
	\end{array}
	\right)
	\quad
	p=
	\left(
	\begin{array}{ccc}
		3/2 \\
		p \\
	\end{array}
	\right)
\end{equation}
\begin{equation}\label{eq:sym133_2}
	S_3(p=gen) = 
	\left(
	\begin{array}{ccc}
		x & y & \bar{z}+1 \\
		x & y & z \\
	\end{array}
	\right)
\end{equation}
$\mathbf{T_1}=(k(2, -2, 2), l(-2, 2, 2), m(2, 2, -2)))$\\
$\mathbf{T_2}=\mathbf{T_3}=(k(2, -2, 2), l(-2, 2, 2)))$\\
The high-symmetric positions $p=\dots,-\frac{1}{2},\frac{3}{2},\frac{7}{2},\dots$ correspond again to the non-polar APBs. 

\section{Free energy invariants and profiles}\label{sec:freenergy}
The macroscopic properties of DWs can be described by the Landau-Ginzburg free energy and it is interesting to compare it with the layer-group approach discussed in the previous section.   
The polarization in the DWs is described by the free energy invariants, which couple the polarization and OP as well as gradient terms. In general there is a large number of invariants and they are shown in the Appendix, see \ref{sec:appendix}. 

\begin{figure} 
	\centering
	\includegraphics[scale=0.4]{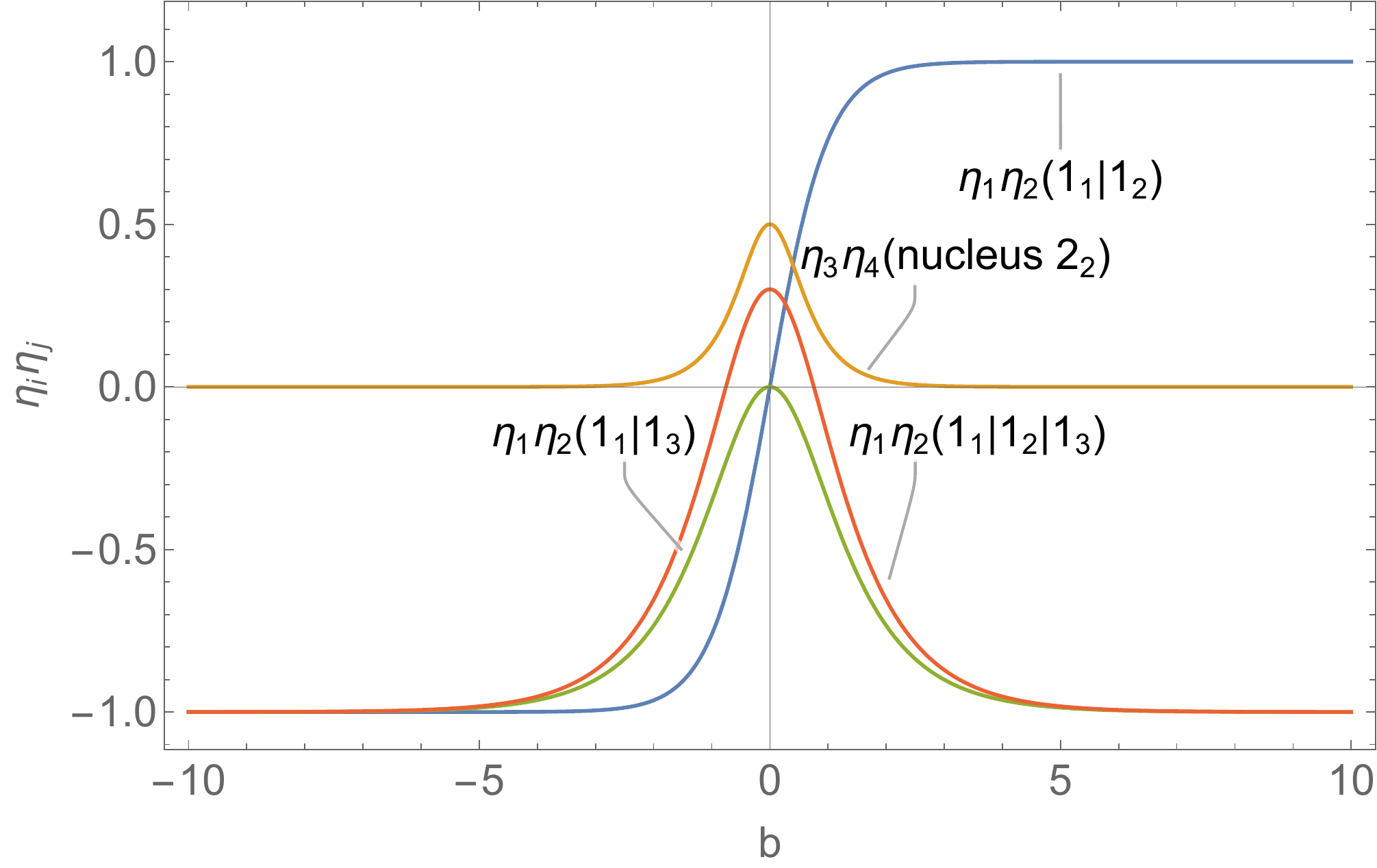}%{fig_profiles.pdf}
	\caption{\textbf{The schematic APB's $\eta_1\eta_2$ profiles.}
		In the APBs with the nucleus $2_2$ the profiles consist of 2 curves: $\eta_1\eta_2$ and $\eta_3\eta_4$. All curves are either even or odd functions. }
	\label{fig:APB_profiles}
\end{figure}

Tagantsev~et.al. \cite{Wei2014} proposed a biquadratic coupling of polarization and OP to explain the polarization in APBs. However, it is clear that the biquadratic coupling terms alone cannot explain the wealth of observed polarization profiles. Here we focus on bilinear couplings of polarization and gradients of OP. The biquadratic and bilinear cases will be compared in terms of their implications to explain the observations. The APBs in the previous sections have the DW normal $\mathbf{n}\perp z$ and the only nonzero OP components $\eta_1,\dots,\eta_6$. Therefore, it is enough to assume $\partial_z(\cdot)=0$ and $\eta_7=\dots=\eta_{12}=0$. If we further consider only '1' and '2' orientational states then also $\eta_5=\eta_6=0$ and the number of nonzero invariants reduces to 12. Only the following 2 of them give the in-plane polarization $\mathbf{P\perp n}=0$:
\begin{eqnarray}\label{eq:U}
	U_{L5} &=&
	(P_x \partial_x-P_y \partial_y)\left(\eta _1 \eta _2 \eta _3 \eta _4\right)=
	(P_a \partial_b+P_b \partial_a)\left(\eta _1 \eta _2 \eta _3 \eta _4\right)=
	P_a \partial_b\left(\eta _1 \eta _2 \eta _3 \eta _4\right)
	 \\ \label{eq:V}
	V_{T1} &=&
	\eta _3 \eta _4 (P_y\partial_x-P_x\partial_y)\left(\eta _1 \eta _2\right)+ \eta _1 \eta _2 (P_x\partial_y-P_y\partial_x)\left(\eta _3 \eta _4\right) = \nonumber \\
	&=&
	\eta _1 \eta _2(P_a\partial_b-P_b\partial_a)\left(\eta _3 \eta _4\right) - 
	\eta _3 \eta _4(P_a\partial_b-P_b\partial_a)\left(\eta _1 \eta _2\right) = \nonumber \\
	&=& P_a(\eta _1 \eta _2\partial_b\left(\eta _3 \eta _4\right) - 
	\eta _3 \eta _4\partial_b\left(\eta _1 \eta _2\right))
\end{eqnarray}
where the orthorhombic axis $b$ is along $\mathbf{n}=(1,1,0)$ and $a$ is along $(1,-1,0)$, and  
$b=\frac{1}{\sqrt{2}}(x+y)$, $a=\frac{1}{\sqrt{2}}(x-y)$, see also insets in Figs.~\ref{fig:1112}, \ref{fig:112112}. The simplest profiles of quadratic terms  $\eta_i\eta_j$ satisfying the boundary conditions of individual APBs are sketched in Fig.~\ref{fig:APB_profiles}. The polarization profiles of APBs can be obtained from the invariants (\ref{eq:U}), (\ref{eq:V}) taking into account that the quadratic forms $\eta_i\eta_j$ are either odd or even functions as shown in Fig.~\ref{fig:APB_profiles},  e.~g., $\eta_1\eta_2$ is an odd profile in $(1_1|1_2)$ APB, while it is an even function in $(1_1|1_3)$ APB. 
\begin{figure} 
	\centering
	\includegraphics[scale=0.4]{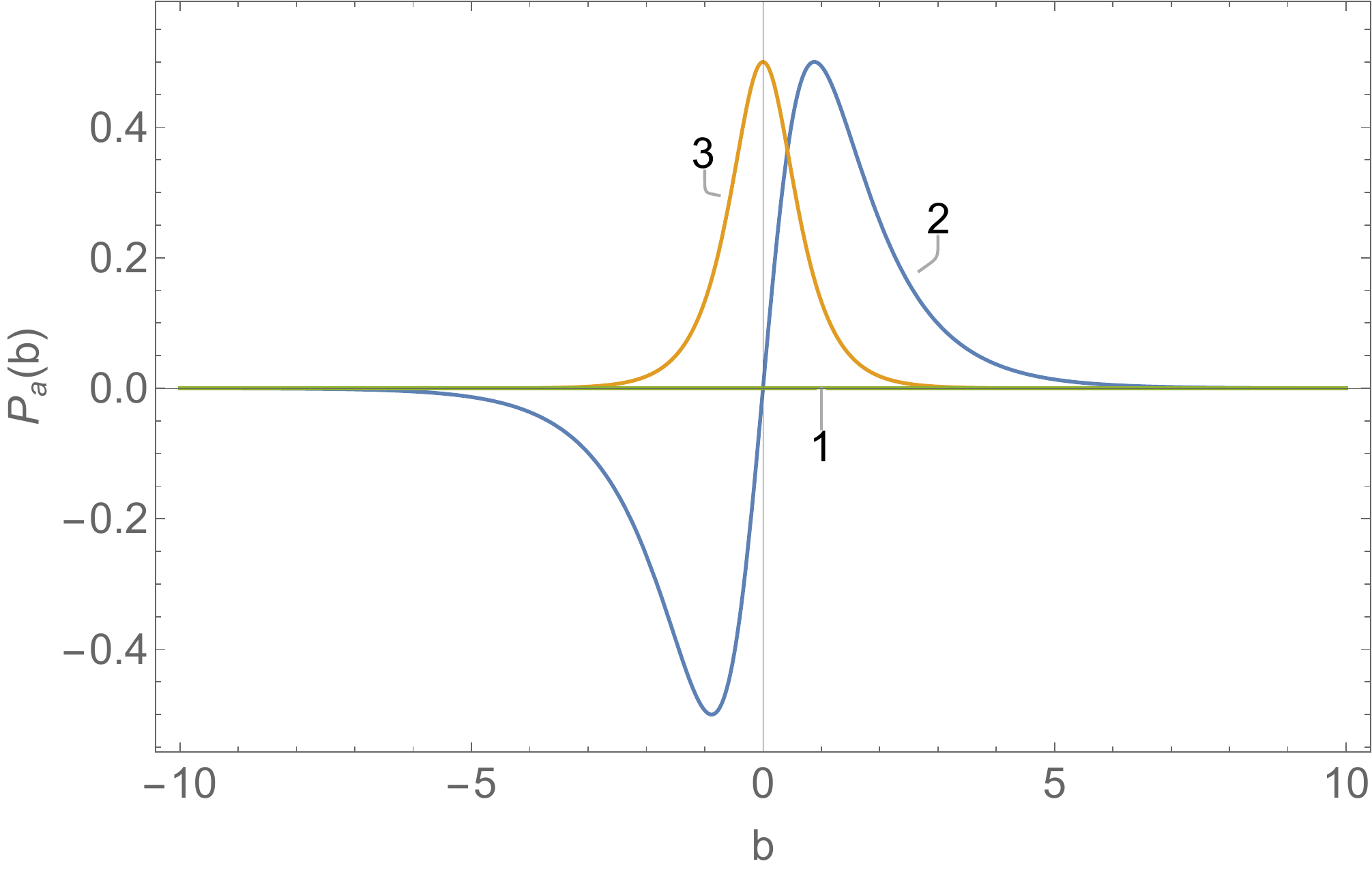}%{fig_profiles_P.pdf}
	\caption{\textbf{The schematic polarization $P_a(b)$ profiles.}
		1: APBs $(1_1|1_2)$, $(1_1|1_3)$, and $(1_1|1_2|1_3)$; 2: APB $(1_1|2_2|1_3)$; 3: APB $(1_1|2_2|1_2)$.  }
	\label{fig:APB_profiles_P}
\end{figure}

Their properties follow also from the symmetries $S_2$. Using the symmetry operations $S_2$ in (\ref{eq:sym1}), (\ref{eq:sym13_1}), the profiles of the APBs $(1_1|1_2)$, $(1_1|1_3)$, and $(1_1|1_2|1_3)$ satisfy $P_a(b)=-P_a(-b)=P_a(-b)\equiv 0$, in agreement with the curve '1' in Fig.~\ref{fig:APB_profiles_P}. The polarization profile of $(1_1|2_2|1_3)$ according to $S_2$ in (\ref{eq:sym123_1}) is asymmetric (odd): $P_a(b)=-P_a(-b)$, as shown in curve '2'. Finally, based on $S_2$ in (\ref{eq:sym3}), the APB $(1_1|2_2|1_2)$ has symmetrical(even) profile since: $P_a(b)=P_a(-b)$ (see curve '3'), i.~e. the only case with macroscopic in-plane polarization. 

\section{Discussion}
The symmetry and properties of DWs on the microscopic level, i.~e. in the sense that the DW's position inside the unit cell is also important can be well described by layer groups \cite{Janovec2006,Schranz2019,Schranz2020,Schranz2020(2)}. Here this method was systematically applied to PZO by using the irreps of OPs to calculate the symmetries of the planar APBs. 
This method unveils that the microscopic polar properties of APBs depend on the APB type and their position. The APBs composed of the translational domains $1_i$ only, i.~e. $(1_1|1_2)$, $(1_1|1_3)$, $(1_1|1_2|1_3)$ and $(1_1|1_4|1_3)$, possess 3 high-symmetry positions, 2 of them at $p=1/2\ \textrm{and}\ 5/2$ are polar with opposite polarizations, while the central position of higher symmetry in-between has antisymmetric polarization profile, see Fig.~\ref{fig:1112}, \ref{fig:1113}. Several APBs of this kind were observed and analyzed by Wei, et.~al. \cite{Wei2014,Wei2015} and their properties can be understood using our approach. E.~g., they measured and modelled the APB denoted as III-1 (Fig.~1 in Wei.~et.~al.\cite{Wei2015}) and observed a double peak of polarization. We identify it as the improper N\'eel-type APB $(1_1|1_4|1_3)$ at $p=1/2$ with nonzero polarization of the double peak shape, Fig~\ref{fig:1113}. The double peak arises from 2 interfaces $1_1|1_4$ and $1_4|1_3$ nearby the boundary center. There exists, as mentioned above, an equivalent position of the APB $p=5/2$ with opposite polarization. In agreement with this, the observation of polar $(1_1|1_4|1_3)$ APB at two microscopic positions was reported \cite{Wei2015}. For comparison, APBs I-1, I-2 in Ref.~\cite{Wei2015} are identified here as $(1_1|1_2)$ at 2 positions (Fig.~\ref{fig:1112}), II-1 and II-2 are $(1_1|1_4)$ at 2 positions, and III-2, III-3 are $(1_1|1_3)$ at 2 positions (Fig.~\ref{fig:1113}). The polarization profiles at the right side of Figs.~\ref{fig:1112}, \ref{fig:112112}, \ref{fig:1113} were calculated using the method of 'sliding orthorhombic unit cell' \cite{Wei2014}, but it should be only treated as approximate visualization, e.~g., the average in-plane polarization in the APB $(1_1|2_1|1_2)$ shown in Fig.~\ref{fig:112112} is not 0. The decisive element for the microscopic polar properties is the layer group symmetry $S_3$ of each APB. 

The DWs are also often described by Landau-Ginzburg theory. 
Here we argue that such coarse grained theory cannot explain all features of the APBs, because their properties depend on the microscopic position inside the unit cell, which cannot be fully accounted by the  phenomenological approach. In previous Secs.~\ref{sec:symm-define}--\ref*{sec:11|13} we have demonstrated that the polarization (including its sign) in some APBs is obtained simply by appropriate choice of the APB position inside the unit cell. In general the microscopic properties of APBs are properly described by the layer groups ($S_3$) only, but still there is a connection with the phenomenological description, which is determined by the average symmetry $S_2$. For that we use the following notation: $\bar{P}_a$ is an average in-plane polarization, which is the same for both the layer group description and the phenomenological description, $P_a(b)$ is a macroscopic (Landau-Ginzburg) polarization profile allowed by $S_2$ and described by Eqs.~(\ref{eq:U}), (\ref{eq:V}). 
Let us summarize the behavior of macroscopic in-plane polarization $P_a$ determined by the symmetry $S_2$ for 3 instructive examples:\\
(i) In purely translational APBs $(1_1|1_2)$, $(1_1|1_3)$, $(1_1|1_2|1_3)$: $\bar{P}_a=0$, $P_a(b)\equiv 0$.
(ii) In the APB with 2 different orientational states $(1_1|2_1|1_3)$: $\bar{P}_a=0$, $P_a(b)=-P_a(-b)\neq 0$. 
(iii) In the APB with 2 different orientational states $(1_1|2_1|1_2)$: $\bar{P}_a\neq 0$, $P_a(b)=P_a(-b)\neq 0$. The corresponding macroscopic profiles are plotted in Fig.~\ref{fig:APB_profiles_P}. 
This entitles us to conclude that the macroscopic in-plane polarization (symmetric or antisymmetric profiles) can exist only in the APBs with a different orientational state at the center. It is in agreement with the macroscopic polarization $P_a$  described by the free energy invariants Eqs.~(\ref{eq:U}), (\ref{eq:V}), the latter is sometimes called 'roto-polar'. For comparison, the biquadratic coupling $P_i^2\eta_j^2$ proposed in Ref.~\cite{Wei2014} implies $\bar{P}_a\ne 0$ for the APB $(1_1|1_3)$, but it contradicts with point (i). 

The  polarization of the APBs observed in Ref.~\cite{Wei2015}, which contain only translational domain states, cannot be described by the (continuous) Landau-Ginzburg theory, but only by the layer-group approach. It is worth to mention that one can distinguish 2 types of APBs with the nucleus (DS precursor) at the center (Sec.~\ref{sec:symm-define}). The first one is an 'improper' N\'eel-type structure in the sense that the APB contains only translational domain states (e.~g., $(1_1|1_4|1_3)$ displayed as path '2' circumventing the origin in Fig.~\ref{fig:PZO domain states}, in which all Pb displacements are parallel) and its symmetry is identical with $(1_1|1_3)$. The second one is a 'proper' N\'eel-type structure with a different orientational state at the center (e.~g. $(1_1|2_1|1_3)$), in which the symmetry is lowered compared to the APB $(1_1|1_3)$. Such a N\'eel structure can develop by means of the phase transition in the APB and its properties (e.g.,~symmetric in-plane polarization) can be well described by Landau-Ginzburg theory, which are fully in accord with the layer group theory.  

The symmetry properties of the APBs determine possibilities of the (in-plane) polarization  reversal in both types of the APBs by reversing the homogeneous electric field $E_a$ applied along the $a$-axis. Two cases can be distinguished. The reversal of the in-plane microscopic polarization in the macroscopically non-polar APBs is accompanied with the shift of the APB position within the unit cell, e.~g. the polarization $+P$ of the APB $(1_1|1_3)$ at $p=1/2$ is reversed to $-P$ via a shift to $p=5/2$ (or -3/2). The polarization switching in the macroscopically polar APBs is rendered by the transformation of the nucleus (DS precursor) at the center to a new DS, e.~g. $+P$ in the APB $(1_1|2_1|1_2)$ is switched to $-P$ in the APB $(1_1|2_2|1_2)$, as can be also seen from Eq.~(\ref{eq:V}). The switching mechanism is illustrated in Fig.~\ref{fig:112112}. 
   
The method used in this contribution for APBs in PZO is readily applicable for DWs in other materials.   

\section{Acknowledgments}
This work was supported by Operational Program Research, Development and Education (financed by European Structural and Investment Funds and by the Czech Ministry of Education, Youth, and Sports), Project No. SOLID21-CZ.02.1.01/0.0/0.0/16\_019/0000760). 

\newpage

\section{Appendix}\label{sec:appendix}
The occurrence of polarization in the DWs can be described by the free energy invariants with bilinear coupling of polarization and OP gradients. In general there is a huge number of invariants, but it can be simplified considering DWs from previous Section with the DW normal $\mathbf{n}\perp z$ and the only nonzero OP components $\eta_1,\dots,\eta_6$. Therefore, in the following text, we assume $\partial_z=0$ and $\eta_7=\dots=\eta_{12}=0$. 
The are only 3 lowest order invariants with quadratic OP components:
\begin{eqnarray}
	T_1 &=& 
	P_y \partial_x \left(\eta _1^2+\eta _2^2-\eta _3^2-\eta _4^2\right)+P_x \partial_y \left(\eta _1^2+\eta _2^2-\eta _3^2-\eta _4^2\right)+ P_z\partial_x \left(\eta _5^2+\eta _6^2)\right) \\
	T_2 &=&  
P_y \partial_y \left(\eta _5^2+\eta _6^2\right) \\
	T_3 &=& 
P_x \partial_x \left(\eta _1^2+\eta _2^2+\eta _3^2+\eta _4^2+\eta _5^2+\eta _6^2\right)+P_y \partial_y \left(\eta _1^2+\eta _2^2+\eta _3^2+\eta _4^2\right)	
\end{eqnarray}

	\begin{table}[h]
	\begin{equation}
\begin{array}{|c|c|c|c|}
	\hline
	\text{} & \multicolumn{3}{c|}{\text{Number of Invariants}}   \\
	\text{} & \multicolumn{3}{c|}{\text{39} }   \\ \hline
	\text{} & \partial _m\left(\eta _i\eta _j\eta _k\eta _l\right) & \eta _i\eta _j\partial _m\left(\eta _k\eta _l\text{)-}\eta _k\eta _l\partial _m\right(\eta _i\eta _j) & \eta _i\eta _j\partial _m\left(\eta _k\eta _l\right) \\
	\text{} & 20 & 7 & 12 \\ \hline
	P_i \partial_i & 15 & 0 & 6 \\
	P_i \partial_j & 5 & 7 & 6 \\ \hline
\end{array}
	\end{equation}
	\caption{\label{tab:2} Total number of invariants. The invariant forms and their count are indicated. E.~g., in the 1st column: number of invariants of type $P_m\partial_m(\eta_i\eta_j\eta_k\eta_l)$ is 15 (longitudinal); number of invariants of type $P_n\partial_m(\eta_i\eta_j\eta_k\eta_l)$ is 5 (transverse); in total it is 20.}
\end{table}

	\begin{table}[h]
	\begin{equation}
\begin{array}{|c|c|c|c|}
	\hline
	\text{} & \multicolumn{3}{c|}{\text{Number of Invariants} \ (\partial_z=0, \eta_i=0, i=7,\dots,12)}   \\
	\text{} & \multicolumn{3}{c|}{\text{25} }   \\ \hline
	\text{} & \partial _m\left(\eta _i\eta _j\eta _k\eta _l\right) & \eta _i\eta _j\partial _m\left(\eta _k\eta _l\text{)-}\eta _k\eta _l\partial _m\right(\eta _i\eta _j) & \eta _i\eta _j\partial _m\left(\eta _k\eta _l\right) \\
	\text{} & 14 & 3 & 8 \\ \hline
	P_i \partial_i & 9 & 0 & 2 \\
	P_i \partial_j & 5 & 3 & 6 \\ \hline
\end{array}
	\end{equation}
	\caption{\label{tab:3} Total number of invariants under assumptions $\partial_z=0, \eta_i=0, i=7,\dots,12$. The invariant forms and their count are indicated. E.~g., in the 1st column: number of invariants of type $P_m\partial_m(\eta_i\eta_j\eta_k\eta_l)$ is 9 (longitudinal); number of invariants of type $P_n\partial_m(\eta_i\eta_j\eta_k\eta_l)$ is 5 (transverse); in total it is 14.}
\end{table}

The next lowest invariants are those with quartic coupling of the OP components. There are in total 39 invariants, or 25 when using the above assumptions, see Tables~\ref{tab:2} and \ref{tab:3}. Notations $U, V, W$ correspond to columns in Table~\ref{tab:3}; subscripts $L,T$ denote longitudinal and transverse invariant type and correspond to last 2 rows in the Tables. The invariants $V$ are Lifshitz-like. The invariants are:
\begin{eqnarray}
	U_{T1} &=&
	P_z \partial_y\left(\left(\eta _1^2+\eta _2^2-\eta _3^2-\eta _4^2\right) \left(\eta _5^2+\eta _6^2\right)\right) \\
	U_{T2} &=&
	P_y \partial_x\left(\left(\eta _1^2+\eta _2^2-\eta _3^2-\eta _4^2\right) \left(\eta _5^2+\eta _6^2\right)\right)+P_z \partial_x\left(\left(\eta _1^2+\eta _2^2+\eta _3^2+\eta _4^2\right) \left(\eta _5^2+\eta _6^2\right)\right) \\
	U_{T3} &=&
	P_x \partial_y\left(\left(\eta _1^2+\eta _2^2-\eta _3^2-\eta _4^2\right) \left(\eta _5^2+\eta _6^2\right)\right) \\
	U_{T4} &=&
	P_y \partial_x\left(\eta _1^2 \eta _2^2-\eta _3^2 \eta _4^2\right)+P_x \partial_y\left(\eta _1^2 \eta _2^2-\eta _3^2 \eta _4^2\right)+P_z \partial_x\left(\eta _5^2 \eta _6^2\right) \\
	U_{T5} &=&
	P_y \partial_x\left(\eta _1^4+\eta _2^4-\eta _3^4-\eta _4^4\right)+P_x \partial_y\left(\eta _1^4+\eta _2^4-\eta _3^4-\eta _4^4\right)+P_z \partial_x\left(\eta _5^4+\eta _6^4\right)
\end{eqnarray}

\begin{eqnarray}
	U_{L1} &=&
	P_y \partial_y\left(\eta _5^2 \eta _6^2\right) \\
	U_{L2} &=&
	P_y \partial_y\left(\eta _5^4+\eta _6^4\right) \\
	U_{L3} &=&
	P_y \partial_y\left(\left(\eta _1^2+\eta _2^2+\eta _3^2+\eta _4^2\right) \left(\eta _5^2+\eta _6^2\right)\right) \\
	U_{L4} &=&
	P_x \partial_x\left(\left(\eta _1^2+\eta _2^2+\eta _3^2+\eta _4^2\right) \left(\eta _5^2+\eta _6^2\right)\right) \\
	U_{L5} &=&
	P_x \partial_x\left(\eta _1 \eta _2 \eta _3 \eta _4\right)-P_y \partial_y\left(\eta _1 \eta _2 \eta _3 \eta _4\right) \\
	U_{L6} &=&
	P_x \partial_x\left(\eta _2^2 \eta _3^2+\eta _1^2 \eta _4^2\right)+P_y \partial_y\left(\eta _2^2 \eta _3^2+\eta _1^2 \eta _4^2\right) \\
	U_{L7} &=&
	P_x \partial_x\left(\eta _1^2 \eta _3^2+\eta _2^2 \eta _4^2\right)+P_y \partial_y\left(\eta _1^2 \eta _3^2+\eta _2^2 \eta _4^2\right) \\
	U_{L8} &=&
	P_x \partial_x\left(\eta _1^2 \eta _2^2+\eta _3^2 \eta _4^2+\eta _5^2 \eta _6^2\right)+P_y \partial_y\left(\eta _1^2 \eta _2^2+\eta _3^2 \eta _4^2\right) \\
	U_{L9} &=&
	P_x \partial_x\left(\eta _1^4+\eta _2^4+\eta _3^4+\eta _4^4+\eta _5^4+\eta _6^4\right)+P_y \partial_y\left(\eta _1^4+\eta _2^4+\eta _3^4+\eta _4^4\right)
\end{eqnarray}

\begin{eqnarray}
	V_{T1} &=&
	P_y \left(\eta _3 \eta _4 \partial_x\left(\eta _1 \eta _2\right)-\eta _1 \eta _2 \partial_x\left(\eta _3 \eta _4\right)\right)+P_x \left(\eta _1 \eta _2 \partial_y\left(\eta _3 \eta _4\right)-\eta _3 \eta _4 \partial_y\left(\eta _1 \eta _2\right)\right) \\
	V_{T2} &=&
	P_y \left(\eta _1^2 \partial_x\left(\eta _4^2\right)-\eta _4^2 \partial_x\left(\eta _1^2\right)+\eta _2^2 \partial_x\left(\eta _3^2\right)-\eta _3^2 \partial_x\left(\eta _2^2\right)\right)+ \nonumber \\
	&+&P_x \left(\eta _1^2 \partial_y\left(\eta _4^2\right)-\eta _4^2 \partial_y\left(\eta _1^2\right)+\eta _2^2 \partial_y\left(\eta _3^2\right)-\eta _3^2 \partial_y\left(\eta _2^2\right)\right) \\
	V_{T3} &=&
	P_y \left(\eta _1^2 \partial_x\left(\eta _3^2\right)-\eta _3^2 \partial_x\left(\eta _1^2\right)+\eta _2^2 \partial_x\left(\eta _4^2\right)-\eta _4^2 \partial_x\left(\eta _2^2\right)\right)+ \nonumber \\
	&+&P_x \left(\eta _1^2 \partial_y\left(\eta _3^2\right)-\eta _3^2 \partial_y\left(\eta _1^2\right)+\eta _2^2 \partial_y\left(\eta _4^2\right)-\eta _4^2 \partial_y\left(\eta _2^2\right)\right)
\end{eqnarray}

\begin{eqnarray}
	W_{T1} &=&
	P_z \left(\eta _1^2+\eta _2^2-\eta _3^2-\eta _4^2\right) \partial_y\left(\eta _5^2+\eta _6^2\right) \\
	W_{T2} &=&
	P_y \left(\eta _5^2+\eta _6^2\right) \partial_x\left(\eta _1^2+\eta _2^2-\eta _3^2-\eta _4^2\right)+P_z \left(\eta _1^2+\eta _2^2+\eta _3^2+\eta _4^2\right) \partial_x\left(\eta _5^2+\eta _6^2\right) \\
	W_{T3} &=&
	P_z \left(\eta _5^2+\eta _6^2\right) \partial_y\left(\eta _1^2+\eta _2^2-\eta _3^2-\eta _4^2\right) \\
	W_{T4} &=&
	P_x \left(\eta _5^2+\eta _6^2\right) \partial_y\left(\eta _1^2+\eta _2^2-\eta _3^2-\eta _4^2\right) \\
	W_{T5} &=&
	P_z \left(\eta _5^2+\eta _6^2\right) \partial_x\left(\eta _1^2+\eta _2^2+\eta _3^2+\eta _4^2\right)+P_y \left(\eta _1^2+\eta _2^2-\eta _3^2-\eta _4^2\right) \partial_x\left(\eta _5^2+\eta _6^2\right) \\
	W_{T6} &=&
	P_x \left(\eta _1^2+\eta _2^2-\eta _3^2-\eta _4^2\right) \partial_y\left(\eta _5^2+\eta _6^2\right) \\
\end{eqnarray}

\begin{eqnarray}
	W_{L1} &=&
	P_y \left(\eta _5^2+\eta _6^2\right) \partial_y\left(\eta _1^2+\eta _2^2+\eta _3^2+\eta _4^2\right) \\
	W_{L2} &=&
    P_y \left(\eta _1^2+\eta _2^2+\eta _3^2+\eta _4^2\right) \partial_y\left(\eta _5^2+\eta _6^2\right) \\
\end{eqnarray}

%\section{Comparison with experiment}

%\end{widetext}

%\nocite{*}

%\bibliography{help}

\begin{thebibliography}{36}%
\makeatletter
\providecommand \@ifxundefined [1]{%
 \@ifx{#1\undefined}
}%
\providecommand \@ifnum [1]{%
 \ifnum #1\expandafter \@firstoftwo
 \else \expandafter \@secondoftwo
 \fi
}%
\providecommand \@ifx [1]{%
 \ifx #1\expandafter \@firstoftwo
 \else \expandafter \@secondoftwo
 \fi
}%
\providecommand \natexlab [1]{#1}%
\providecommand \enquote  [1]{``#1''}%
\providecommand \bibnamefont  [1]{#1}%
\providecommand \bibfnamefont [1]{#1}%
\providecommand \citenamefont [1]{#1}%
\providecommand \href@noop [0]{\@secondoftwo}%
\providecommand \href [0]{\begingroup \@sanitize@url \@href}%
\providecommand \@href[1]{\@@startlink{#1}\@@href}%
\providecommand \@@href[1]{\endgroup#1\@@endlink}%
\providecommand \@sanitize@url [0]{\catcode `\\12\catcode `\$12\catcode
  `\&12\catcode `\#12\catcode `\^12\catcode `\_12\catcode `\%12\relax}%
\providecommand \@@startlink[1]{}%
\providecommand \@@endlink[0]{}%
\providecommand \url  [0]{\begingroup\@sanitize@url \@url }%
\providecommand \@url [1]{\endgroup\@href {#1}{\urlprefix }}%
\providecommand \urlprefix  [0]{URL }%
\providecommand \Eprint [0]{\href }%
\providecommand \doibase [0]{https://doi.org/}%
\providecommand \selectlanguage [0]{\@gobble}%
\providecommand \bibinfo  [0]{\@secondoftwo}%
\providecommand \bibfield  [0]{\@secondoftwo}%
\providecommand \translation [1]{[#1]}%
\providecommand \BibitemOpen [0]{}%
\providecommand \bibitemStop [0]{}%
\providecommand \bibitemNoStop [0]{.\EOS\space}%
\providecommand \EOS [0]{\spacefactor3000\relax}%
\providecommand \BibitemShut  [1]{\csname bibitem#1\endcsname}%
\let\auto@bib@innerbib\@empty
%</preamble>
\bibitem [{\citenamefont {Wei}\ \emph {et~al.}(2014)\citenamefont {Wei},
  \citenamefont {Tagantsev}, \citenamefont {Kvasov}, \citenamefont {Roleder},
  \citenamefont {Jia},\ and\ \citenamefont {Setter}}]{Wei2014}%
  \BibitemOpen
  \bibfield  {author} {\bibinfo {author} {\bibfnamefont {X.-K.}\ \bibnamefont
  {Wei}}, \bibinfo {author} {\bibfnamefont {A.~K.}\ \bibnamefont {Tagantsev}},
  \bibinfo {author} {\bibfnamefont {A.}~\bibnamefont {Kvasov}}, \bibinfo
  {author} {\bibfnamefont {K.}~\bibnamefont {Roleder}}, \bibinfo {author}
  {\bibfnamefont {C.-L.}\ \bibnamefont {Jia}},\ and\ \bibinfo {author}
  {\bibfnamefont {N.}~\bibnamefont {Setter}},\ }\bibfield  {title} {\bibinfo
  {title} {Ferroelectric translational antiphase boundaries in nonpolar
  materials},\ }\href@noop {} {\bibfield  {journal} {\bibinfo  {journal}
  {Nature Communications}\ }\textbf {\bibinfo {volume} {5}},\ \bibinfo {pages}
  {3031} (\bibinfo {year} {2014})}\BibitemShut {NoStop}%
\bibitem [{\citenamefont {Wei}\ \emph {et~al.}(2015)\citenamefont {Wei},
  \citenamefont {Jia}, \citenamefont {Roleder},\ and\ \citenamefont
  {Setter}}]{Wei2015}%
  \BibitemOpen
  \bibfield  {author} {\bibinfo {author} {\bibfnamefont {X.-K.}\ \bibnamefont
  {Wei}}, \bibinfo {author} {\bibfnamefont {C.-L.}\ \bibnamefont {Jia}},
  \bibinfo {author} {\bibfnamefont {K.}~\bibnamefont {Roleder}},\ and\ \bibinfo
  {author} {\bibfnamefont {N.}~\bibnamefont {Setter}},\ }\bibfield  {title}
  {\bibinfo {title} {Polarity of translation boundaries in antiferroelectric
  {PbZrO}$_3$},\ }\href@noop {} {\bibfield  {journal} {\bibinfo  {journal}
  {Materials research bulletin}\ }\textbf {\bibinfo {volume} {62}},\ \bibinfo
  {pages} {101} (\bibinfo {year} {2015})}\BibitemShut {NoStop}%
\bibitem [{\citenamefont {Van~Aert}\ \emph {et~al.}(2012)\citenamefont
  {Van~Aert}, \citenamefont {Turner}, \citenamefont {Delville}, \citenamefont
  {Schryvers}, \citenamefont {Van~Tendeloo},\ and\ \citenamefont
  {Salje}}]{Aert2012}%
  \BibitemOpen
  \bibfield  {author} {\bibinfo {author} {\bibfnamefont {S.}~\bibnamefont
  {Van~Aert}}, \bibinfo {author} {\bibfnamefont {S.}~\bibnamefont {Turner}},
  \bibinfo {author} {\bibfnamefont {R.}~\bibnamefont {Delville}}, \bibinfo
  {author} {\bibfnamefont {D.}~\bibnamefont {Schryvers}}, \bibinfo {author}
  {\bibfnamefont {G.}~\bibnamefont {Van~Tendeloo}},\ and\ \bibinfo {author}
  {\bibfnamefont {E.~K.}\ \bibnamefont {Salje}},\ }\bibfield  {title} {\bibinfo
  {title} {Direct observation of ferrielectricity at ferroelastic domain
  boundaries in {CaTiO}$_3$ by electron microscopy},\ }\href@noop {} {\bibfield
   {journal} {\bibinfo  {journal} {Advanced Materials}\ }\textbf {\bibinfo
  {volume} {24}},\ \bibinfo {pages} {523} (\bibinfo {year} {2012})}\BibitemShut
  {NoStop}%
\bibitem [{\citenamefont {Yokota}\ \emph {et~al.}(2014)\citenamefont {Yokota},
  \citenamefont {Usami}, \citenamefont {Haumont}, \citenamefont {Hicher},
  \citenamefont {Kaneshiro}, \citenamefont {Salje},\ and\ \citenamefont
  {Uesu}}]{Yokota2014}%
  \BibitemOpen
  \bibfield  {author} {\bibinfo {author} {\bibfnamefont {H.}~\bibnamefont
  {Yokota}}, \bibinfo {author} {\bibfnamefont {H.}~\bibnamefont {Usami}},
  \bibinfo {author} {\bibfnamefont {R.}~\bibnamefont {Haumont}}, \bibinfo
  {author} {\bibfnamefont {P.}~\bibnamefont {Hicher}}, \bibinfo {author}
  {\bibfnamefont {J.}~\bibnamefont {Kaneshiro}}, \bibinfo {author}
  {\bibfnamefont {E.}~\bibnamefont {Salje}},\ and\ \bibinfo {author}
  {\bibfnamefont {Y.}~\bibnamefont {Uesu}},\ }\bibfield  {title} {\bibinfo
  {title} {Direct evidence of polar nature of ferroelastic twin boundaries in
  {CaTiO}$_3$ obtained by second harmonic generation microscope},\ }\href@noop
  {} {\bibfield  {journal} {\bibinfo  {journal} {Physical Review B}\ }\textbf
  {\bibinfo {volume} {89}},\ \bibinfo {pages} {144109} (\bibinfo {year}
  {2014})}\BibitemShut {NoStop}%
\bibitem [{\citenamefont {Yokota}\ \emph {et~al.}(2017)\citenamefont {Yokota},
  \citenamefont {Niki}, \citenamefont {Haumont}, \citenamefont {Hicher},\ and\
  \citenamefont {Uesu}}]{Yokota2017}%
  \BibitemOpen
  \bibfield  {author} {\bibinfo {author} {\bibfnamefont {H.}~\bibnamefont
  {Yokota}}, \bibinfo {author} {\bibfnamefont {S.}~\bibnamefont {Niki}},
  \bibinfo {author} {\bibfnamefont {R.}~\bibnamefont {Haumont}}, \bibinfo
  {author} {\bibfnamefont {P.}~\bibnamefont {Hicher}},\ and\ \bibinfo {author}
  {\bibfnamefont {Y.}~\bibnamefont {Uesu}},\ }\bibfield  {title} {\bibinfo
  {title} {Polar nature of stress-induced twin walls in ferroelastic
  {CaTiO}$_3$},\ }\href@noop {} {\bibfield  {journal} {\bibinfo  {journal} {AIP
  Advances}\ }\textbf {\bibinfo {volume} {7}},\ \bibinfo {pages} {085315}
  (\bibinfo {year} {2017})}\BibitemShut {NoStop}%
\bibitem [{\citenamefont {Salje}\ \emph {et~al.}(2013)\citenamefont {Salje},
  \citenamefont {Aktas}, \citenamefont {Carpenter}, \citenamefont {Laguta},\
  and\ \citenamefont {Scott}}]{Salje2013}%
  \BibitemOpen
  \bibfield  {author} {\bibinfo {author} {\bibfnamefont {E.}~\bibnamefont
  {Salje}}, \bibinfo {author} {\bibfnamefont {O.}~\bibnamefont {Aktas}},
  \bibinfo {author} {\bibfnamefont {M.}~\bibnamefont {Carpenter}}, \bibinfo
  {author} {\bibfnamefont {V.}~\bibnamefont {Laguta}},\ and\ \bibinfo {author}
  {\bibfnamefont {J.}~\bibnamefont {Scott}},\ }\bibfield  {title} {\bibinfo
  {title} {Domains within domains and walls within walls: Evidence for polar
  domains in cryogenic {SrTiO}$_3$},\ }\href@noop {} {\bibfield  {journal}
  {\bibinfo  {journal} {Physical Review Letters}\ }\textbf {\bibinfo {volume}
  {111}},\ \bibinfo {pages} {247603} (\bibinfo {year} {2013})}\BibitemShut
  {NoStop}%
\bibitem [{\citenamefont {Lei}\ \emph {et~al.}(2018)\citenamefont {Lei},
  \citenamefont {Fan}, \citenamefont {Fang}, \citenamefont {Ren}, \citenamefont
  {Ma},\ and\ \citenamefont {Tian}}]{Lei2018}%
  \BibitemOpen
  \bibfield  {author} {\bibinfo {author} {\bibfnamefont {S.}~\bibnamefont
  {Lei}}, \bibinfo {author} {\bibfnamefont {H.}~\bibnamefont {Fan}}, \bibinfo
  {author} {\bibfnamefont {J.}~\bibnamefont {Fang}}, \bibinfo {author}
  {\bibfnamefont {X.}~\bibnamefont {Ren}}, \bibinfo {author} {\bibfnamefont
  {L.}~\bibnamefont {Ma}},\ and\ \bibinfo {author} {\bibfnamefont
  {H.}~\bibnamefont {Tian}},\ }\bibfield  {title} {\bibinfo {title} {Unusual
  devisable high-performance perovskite materials obtained by engineering in
  twins, domains, and antiphase boundaries},\ }\href@noop {} {\bibfield
  {journal} {\bibinfo  {journal} {Inorganic Chemistry Frontiers}\ }\textbf
  {\bibinfo {volume} {5}},\ \bibinfo {pages} {568} (\bibinfo {year}
  {2018})}\BibitemShut {NoStop}%
\bibitem [{\citenamefont {Nataf}\ \emph {et~al.}(2020)\citenamefont {Nataf},
  \citenamefont {Guennou}, \citenamefont {Gregg}, \citenamefont {Meier},
  \citenamefont {Hlinka}, \citenamefont {Salje},\ and\ \citenamefont
  {Kreisel}}]{Nataf2020}%
  \BibitemOpen
  \bibfield  {author} {\bibinfo {author} {\bibfnamefont {G.}~\bibnamefont
  {Nataf}}, \bibinfo {author} {\bibfnamefont {M.}~\bibnamefont {Guennou}},
  \bibinfo {author} {\bibfnamefont {J.}~\bibnamefont {Gregg}}, \bibinfo
  {author} {\bibfnamefont {D.}~\bibnamefont {Meier}}, \bibinfo {author}
  {\bibfnamefont {J.}~\bibnamefont {Hlinka}}, \bibinfo {author} {\bibfnamefont
  {E.}~\bibnamefont {Salje}},\ and\ \bibinfo {author} {\bibfnamefont
  {J.}~\bibnamefont {Kreisel}},\ }\bibfield  {title} {\bibinfo {title}
  {Domain-wall engineering and topological defects in ferroelectric and
  ferroelastic materials},\ }\href@noop {} {\bibfield  {journal} {\bibinfo
  {journal} {Nature Reviews Physics}\ }\textbf {\bibinfo {volume} {2}},\
  \bibinfo {pages} {634} (\bibinfo {year} {2020})}\BibitemShut {NoStop}%
\bibitem [{\citenamefont {Catalan}\ \emph {et~al.}(2012)\citenamefont
  {Catalan}, \citenamefont {Seidel}, \citenamefont {Ramesh},\ and\
  \citenamefont {Scott}}]{Catalan2012}%
  \BibitemOpen
  \bibfield  {author} {\bibinfo {author} {\bibfnamefont {G.}~\bibnamefont
  {Catalan}}, \bibinfo {author} {\bibfnamefont {J.}~\bibnamefont {Seidel}},
  \bibinfo {author} {\bibfnamefont {R.}~\bibnamefont {Ramesh}},\ and\ \bibinfo
  {author} {\bibfnamefont {J.~F.}\ \bibnamefont {Scott}},\ }\bibfield  {title}
  {\bibinfo {title} {Domain wall nanoelectronics},\ }\href@noop {} {\bibfield
  {journal} {\bibinfo  {journal} {Reviews of Modern Physics}\ }\textbf
  {\bibinfo {volume} {84}},\ \bibinfo {pages} {119} (\bibinfo {year}
  {2012})}\BibitemShut {NoStop}%
\bibitem [{\citenamefont {Salje}\ \emph {et~al.}(2016)\citenamefont {Salje},
  \citenamefont {Alexe}, \citenamefont {Kustov}, \citenamefont {Weber},
  \citenamefont {Schiemer}, \citenamefont {Nataf},\ and\ \citenamefont
  {Kreisel}}]{Salje2016}%
  \BibitemOpen
  \bibfield  {author} {\bibinfo {author} {\bibfnamefont {E.~K.}\ \bibnamefont
  {Salje}}, \bibinfo {author} {\bibfnamefont {M.}~\bibnamefont {Alexe}},
  \bibinfo {author} {\bibfnamefont {S.}~\bibnamefont {Kustov}}, \bibinfo
  {author} {\bibfnamefont {M.~C.}\ \bibnamefont {Weber}}, \bibinfo {author}
  {\bibfnamefont {J.}~\bibnamefont {Schiemer}}, \bibinfo {author}
  {\bibfnamefont {G.~F.}\ \bibnamefont {Nataf}},\ and\ \bibinfo {author}
  {\bibfnamefont {J.}~\bibnamefont {Kreisel}},\ }\bibfield  {title} {\bibinfo
  {title} {Direct observation of polar tweed in {LaAlO}$_3$},\ }\href@noop {}
  {\bibfield  {journal} {\bibinfo  {journal} {Scientific reports}\ }\textbf
  {\bibinfo {volume} {6}},\ \bibinfo {pages} {1} (\bibinfo {year}
  {2016})}\BibitemShut {NoStop}%
\bibitem [{\citenamefont {Yokota}\ \emph {et~al.}(2020)\citenamefont {Yokota},
  \citenamefont {Hasegawa}, \citenamefont {Glazer}, \citenamefont {Salje},\
  and\ \citenamefont {Uesu}}]{Salje2020}%
  \BibitemOpen
  \bibfield  {author} {\bibinfo {author} {\bibfnamefont {H.}~\bibnamefont
  {Yokota}}, \bibinfo {author} {\bibfnamefont {N.}~\bibnamefont {Hasegawa}},
  \bibinfo {author} {\bibfnamefont {M.}~\bibnamefont {Glazer}}, \bibinfo
  {author} {\bibfnamefont {E.}~\bibnamefont {Salje}},\ and\ \bibinfo {author}
  {\bibfnamefont {Y.}~\bibnamefont {Uesu}},\ }\bibfield  {title} {\bibinfo
  {title} {Direct evidence of polar ferroelastic domain boundaries in
  semiconductor {BiVO}$_4$},\ }\href@noop {} {\bibfield  {journal} {\bibinfo
  {journal} {Applied Physics Letters}\ }\textbf {\bibinfo {volume} {116}},\
  \bibinfo {pages} {232901} (\bibinfo {year} {2020})}\BibitemShut {NoStop}%
\bibitem [{\citenamefont {Gu}\ \emph {et~al.}(2014)\citenamefont {Gu},
  \citenamefont {Li}, \citenamefont {Morozovska}, \citenamefont {Wang},
  \citenamefont {Eliseev}, \citenamefont {Gopalan},\ and\ \citenamefont
  {Chen}}]{Gu2014}%
  \BibitemOpen
  \bibfield  {author} {\bibinfo {author} {\bibfnamefont {Y.}~\bibnamefont
  {Gu}}, \bibinfo {author} {\bibfnamefont {M.}~\bibnamefont {Li}}, \bibinfo
  {author} {\bibfnamefont {A.~N.}\ \bibnamefont {Morozovska}}, \bibinfo
  {author} {\bibfnamefont {Y.}~\bibnamefont {Wang}}, \bibinfo {author}
  {\bibfnamefont {E.~A.}\ \bibnamefont {Eliseev}}, \bibinfo {author}
  {\bibfnamefont {V.}~\bibnamefont {Gopalan}},\ and\ \bibinfo {author}
  {\bibfnamefont {L.-Q.}\ \bibnamefont {Chen}},\ }\bibfield  {title} {\bibinfo
  {title} {Flexoelectricity and ferroelectric domain wall structures:
  Phase-field modeling and {DFT} calculations},\ }\href@noop {} {\bibfield
  {journal} {\bibinfo  {journal} {Physical Review B}\ }\textbf {\bibinfo
  {volume} {89}},\ \bibinfo {pages} {174111} (\bibinfo {year}
  {2014})}\BibitemShut {NoStop}%
\bibitem [{\citenamefont {Schiaffino}\ and\ \citenamefont
  {Stengel}(2017)}]{Stengel2017}%
  \BibitemOpen
  \bibfield  {author} {\bibinfo {author} {\bibfnamefont {A.}~\bibnamefont
  {Schiaffino}}\ and\ \bibinfo {author} {\bibfnamefont {M.}~\bibnamefont
  {Stengel}},\ }\bibfield  {title} {\bibinfo {title} {Macroscopic polarization
  from antiferrodistortive cycloids in ferroelastic {SrTiO}$_3$},\ }\href@noop
  {} {\bibfield  {journal} {\bibinfo  {journal} {Physical Review Letters}\
  }\textbf {\bibinfo {volume} {119}},\ \bibinfo {pages} {137601} (\bibinfo
  {year} {2017})}\BibitemShut {NoStop}%
\bibitem [{\citenamefont {Janovec}\ and\ \citenamefont
  {P{\v{r}}{\'{\i}}vratsk{\'{a}}}(2006)}]{Janovec2006}%
  \BibitemOpen
  \bibfield  {author} {\bibinfo {author} {\bibfnamefont {V.}~\bibnamefont
  {Janovec}}\ and\ \bibinfo {author} {\bibfnamefont {J.}~\bibnamefont
  {P{\v{r}}{\'{\i}}vratsk{\'{a}}}},\ }\bibfield  {title} {\bibinfo {title}
  {Domain structures},\ }in\ \href
  {https://doi.org/10.1107/97809553602060000645} {\emph {\bibinfo {booktitle}
  {International Tables for Crystallography}}}\ (\bibinfo  {publisher}
  {International Union of Crystallography},\ \bibinfo {year} {2006})\ pp.\
  \bibinfo {pages} {449--505}\BibitemShut {NoStop}%
\bibitem [{\citenamefont {Janovec}\ \emph {et~al.}(1989)\citenamefont
  {Janovec}, \citenamefont {Schranz}, \citenamefont {Warhanek},\ and\
  \citenamefont {Zikmund}}]{Janovec1989}%
  \BibitemOpen
  \bibfield  {author} {\bibinfo {author} {\bibfnamefont {V.}~\bibnamefont
  {Janovec}}, \bibinfo {author} {\bibfnamefont {W.}~\bibnamefont {Schranz}},
  \bibinfo {author} {\bibfnamefont {H.}~\bibnamefont {Warhanek}},\ and\
  \bibinfo {author} {\bibfnamefont {Z.}~\bibnamefont {Zikmund}},\ }\bibfield
  {title} {\bibinfo {title} {Symmetry analysis of domain structure in {KSCN}
  crystals},\ }\href {https://doi.org/10.1080/00150198908217581} {\bibfield
  {journal} {\bibinfo  {journal} {Ferroelectrics}\ }\textbf {\bibinfo {volume}
  {98}},\ \bibinfo {pages} {171} (\bibinfo {year} {1989})}\BibitemShut
  {NoStop}%
\bibitem [{\citenamefont {Ishibashi}\ and\ \citenamefont
  {Dvořák}(1976)}]{Ishibashi1976}%
  \BibitemOpen
  \bibfield  {author} {\bibinfo {author} {\bibfnamefont {Y.}~\bibnamefont
  {Ishibashi}}\ and\ \bibinfo {author} {\bibfnamefont {V.}~\bibnamefont
  {Dvořák}},\ }\bibfield  {title} {\bibinfo {title} {Domain walls in improper
  ferroelectrics},\ }\href {https://doi.org/10.1143/JPSJ.41.1650} {\bibfield
  {journal} {\bibinfo  {journal} {Journal of the Physical Society of Japan}\
  }\textbf {\bibinfo {volume} {41}},\ \bibinfo {pages} {1650} (\bibinfo {year}
  {1976})},\ \Eprint
  {https://arxiv.org/abs/https://doi.org/10.1143/JPSJ.41.1650}
  {https://doi.org/10.1143/JPSJ.41.1650} \BibitemShut {NoStop}%
\bibitem [{\citenamefont {Bullbich}\ and\ \citenamefont
  {Gufan}(1989)}]{Bullbich1989}%
  \BibitemOpen
  \bibfield  {author} {\bibinfo {author} {\bibfnamefont {A.~A.}\ \bibnamefont
  {Bullbich}}\ and\ \bibinfo {author} {\bibfnamefont {Y.~M.}\ \bibnamefont
  {Gufan}},\ }\bibfield  {title} {\bibinfo {title} {Phase transitions in domain
  walls},\ }\href {https://doi.org/10.1080/00150198908217589} {\bibfield
  {journal} {\bibinfo  {journal} {Ferroelectrics}\ }\textbf {\bibinfo {volume}
  {98}},\ \bibinfo {pages} {277} (\bibinfo {year} {1989})},\ \Eprint
  {https://arxiv.org/abs/https://doi.org/10.1080/00150198908217589}
  {https://doi.org/10.1080/00150198908217589} \BibitemShut {NoStop}%
\bibitem [{\citenamefont {Houchmandzadeh}\ \emph {et~al.}(1991)\citenamefont
  {Houchmandzadeh}, \citenamefont {Lajzerowicz},\ and\ \citenamefont
  {Salje}}]{Salje1991}%
  \BibitemOpen
  \bibfield  {author} {\bibinfo {author} {\bibfnamefont {B.}~\bibnamefont
  {Houchmandzadeh}}, \bibinfo {author} {\bibfnamefont {J.}~\bibnamefont
  {Lajzerowicz}},\ and\ \bibinfo {author} {\bibfnamefont {E.}~\bibnamefont
  {Salje}},\ }\bibfield  {title} {\bibinfo {title} {Order parameter coupling
  and chirality of domain walls},\ }\href
  {https://doi.org/10.1088/0953-8984/3/27/009} {\bibfield  {journal} {\bibinfo
  {journal} {Journal of Physics: Condensed Matter}\ }\textbf {\bibinfo {volume}
  {3}},\ \bibinfo {pages} {5163} (\bibinfo {year} {1991})}\BibitemShut
  {NoStop}%
\bibitem [{\citenamefont {Tagantsev}\ \emph {et~al.}(2001)\citenamefont
  {Tagantsev}, \citenamefont {Courtens},\ and\ \citenamefont
  {Arzel}}]{Tagantsev2001}%
  \BibitemOpen
  \bibfield  {author} {\bibinfo {author} {\bibfnamefont {A.~K.}\ \bibnamefont
  {Tagantsev}}, \bibinfo {author} {\bibfnamefont {E.}~\bibnamefont
  {Courtens}},\ and\ \bibinfo {author} {\bibfnamefont {L.}~\bibnamefont
  {Arzel}},\ }\bibfield  {title} {\bibinfo {title} {Prediction of a
  low-temperature ferroelectric instability in antiphase domain boundaries of
  strontium titanate},\ }\href {https://doi.org/10.1103/PhysRevB.64.224107}
  {\bibfield  {journal} {\bibinfo  {journal} {Phys. Rev. B}\ }\textbf {\bibinfo
  {volume} {64}},\ \bibinfo {pages} {224107} (\bibinfo {year}
  {2001})}\BibitemShut {NoStop}%
\bibitem [{\citenamefont {Sonin}\ and\ \citenamefont
  {Tagantsev}(1989)}]{Sonin1989}%
  \BibitemOpen
  \bibfield  {author} {\bibinfo {author} {\bibfnamefont {E.~B.}\ \bibnamefont
  {Sonin}}\ and\ \bibinfo {author} {\bibfnamefont {A.~K.}\ \bibnamefont
  {Tagantsev}},\ }\bibfield  {title} {\bibinfo {title} {Structure and phase
  transitions in antiphase boundaries of improper ferroelectrics},\ }\href
  {https://doi.org/10.1080/00150198908217590} {\bibfield  {journal} {\bibinfo
  {journal} {Ferroelectrics}\ }\textbf {\bibinfo {volume} {98}},\ \bibinfo
  {pages} {291} (\bibinfo {year} {1989})},\ \Eprint
  {https://arxiv.org/abs/https://doi.org/10.1080/00150198908217590}
  {https://doi.org/10.1080/00150198908217590} \BibitemShut {NoStop}%
\bibitem [{\citenamefont {Morozovska}\ \emph {et~al.}(2012)\citenamefont
  {Morozovska}, \citenamefont {Eliseev}, \citenamefont {Glinchuk},
  \citenamefont {Chen},\ and\ \citenamefont {Gopalan}}]{Morozovska2012}%
  \BibitemOpen
  \bibfield  {author} {\bibinfo {author} {\bibfnamefont {A.~N.}\ \bibnamefont
  {Morozovska}}, \bibinfo {author} {\bibfnamefont {E.~A.}\ \bibnamefont
  {Eliseev}}, \bibinfo {author} {\bibfnamefont {M.~D.}\ \bibnamefont
  {Glinchuk}}, \bibinfo {author} {\bibfnamefont {L.-Q.}\ \bibnamefont {Chen}},\
  and\ \bibinfo {author} {\bibfnamefont {V.}~\bibnamefont {Gopalan}},\
  }\bibfield  {title} {\bibinfo {title} {Interfacial polarization and
  pyroelectricity in antiferrodistortive structures induced by a flexoelectric
  effect and rotostriction},\ }\href@noop {} {\bibfield  {journal} {\bibinfo
  {journal} {Physical Review B}\ }\textbf {\bibinfo {volume} {85}},\ \bibinfo
  {pages} {094107} (\bibinfo {year} {2012})}\BibitemShut {NoStop}%
\bibitem [{\citenamefont {Tagantsev}\ \emph {et~al.}(2013)\citenamefont
  {Tagantsev}, \citenamefont {Vaideeswaran}, \citenamefont {Vakhrushev},
  \citenamefont {Filimonov}, \citenamefont {Burkovsky}, \citenamefont
  {Shaganov}, \citenamefont {Andronikova}, \citenamefont {Rudskoy},
  \citenamefont {Baron}, \citenamefont {Uchiyama}, \citenamefont {Chernyshov},
  \citenamefont {Bosak}, \citenamefont {Ujma}, \citenamefont {Roleder},
  \citenamefont {Majchrowski}, \citenamefont {Ko},\ and\ \citenamefont
  {Setter}}]{Tagantsev2013}%
  \BibitemOpen
  \bibfield  {author} {\bibinfo {author} {\bibfnamefont {A.~K.}\ \bibnamefont
  {Tagantsev}}, \bibinfo {author} {\bibfnamefont {K.}~\bibnamefont
  {Vaideeswaran}}, \bibinfo {author} {\bibfnamefont {S.~B.}\ \bibnamefont
  {Vakhrushev}}, \bibinfo {author} {\bibfnamefont {A.~V.}\ \bibnamefont
  {Filimonov}}, \bibinfo {author} {\bibfnamefont {R.~G.}\ \bibnamefont
  {Burkovsky}}, \bibinfo {author} {\bibfnamefont {A.}~\bibnamefont {Shaganov}},
  \bibinfo {author} {\bibfnamefont {D.}~\bibnamefont {Andronikova}}, \bibinfo
  {author} {\bibfnamefont {A.~I.}\ \bibnamefont {Rudskoy}}, \bibinfo {author}
  {\bibfnamefont {A.~Q.~R.}\ \bibnamefont {Baron}}, \bibinfo {author}
  {\bibfnamefont {H.}~\bibnamefont {Uchiyama}}, \bibinfo {author}
  {\bibfnamefont {D.}~\bibnamefont {Chernyshov}}, \bibinfo {author}
  {\bibfnamefont {A.}~\bibnamefont {Bosak}}, \bibinfo {author} {\bibfnamefont
  {Z.}~\bibnamefont {Ujma}}, \bibinfo {author} {\bibfnamefont {K.}~\bibnamefont
  {Roleder}}, \bibinfo {author} {\bibfnamefont {A.}~\bibnamefont
  {Majchrowski}}, \bibinfo {author} {\bibfnamefont {J.-H.}\ \bibnamefont
  {Ko}},\ and\ \bibinfo {author} {\bibfnamefont {N.}~\bibnamefont {Setter}},\
  }\bibfield  {title} {\bibinfo {title} {The origin of antiferroelectricity in
  {PbZrO}$_3$},\ }\href {https://doi.org/10.1038/ncomms3229} {\bibfield
  {journal} {\bibinfo  {journal} {Nature Communications}\ }\textbf {\bibinfo
  {volume} {4}},\ \bibinfo {pages} {2229} (\bibinfo {year} {2013})}\BibitemShut
  {NoStop}%
\bibitem [{\citenamefont {Schranz}\ \emph {et~al.}(2019)\citenamefont
  {Schranz}, \citenamefont {Rychetsky},\ and\ \citenamefont
  {Hlinka}}]{Schranz2019}%
  \BibitemOpen
  \bibfield  {author} {\bibinfo {author} {\bibfnamefont {W.}~\bibnamefont
  {Schranz}}, \bibinfo {author} {\bibfnamefont {I.}~\bibnamefont {Rychetsky}},\
  and\ \bibinfo {author} {\bibfnamefont {J.}~\bibnamefont {Hlinka}},\
  }\bibfield  {title} {\bibinfo {title} {Polarity of domain boundaries in
  nonpolar materials derived from order parameter and layer group symmetry},\
  }\href@noop {} {\bibfield  {journal} {\bibinfo  {journal} {Physical Review
  B}\ }\textbf {\bibinfo {volume} {100}},\ \bibinfo {pages} {184105} (\bibinfo
  {year} {2019})}\BibitemShut {NoStop}%
\bibitem [{\citenamefont {Schranz}\ \emph
  {et~al.}(2020{\natexlab{a}})\citenamefont {Schranz}, \citenamefont
  {Schuster}, \citenamefont {Tr{\"o}ster},\ and\ \citenamefont
  {Rychetsky}}]{Schranz2020}%
  \BibitemOpen
  \bibfield  {author} {\bibinfo {author} {\bibfnamefont {W.}~\bibnamefont
  {Schranz}}, \bibinfo {author} {\bibfnamefont {C.}~\bibnamefont {Schuster}},
  \bibinfo {author} {\bibfnamefont {A.}~\bibnamefont {Tr{\"o}ster}},\ and\
  \bibinfo {author} {\bibfnamefont {I.}~\bibnamefont {Rychetsky}},\ }\bibfield
  {title} {\bibinfo {title} {Polarization of domain boundaries in {SrTiO}$_3$
  studied by layer group and order-parameter symmetry},\ }\href@noop {}
  {\bibfield  {journal} {\bibinfo  {journal} {Physical Review B}\ }\textbf
  {\bibinfo {volume} {102}},\ \bibinfo {pages} {184101} (\bibinfo {year}
  {2020}{\natexlab{a}})}\BibitemShut {NoStop}%
\bibitem [{\citenamefont {Schranz}\ \emph
  {et~al.}(2020{\natexlab{b}})\citenamefont {Schranz}, \citenamefont
  {Tröster},\ and\ \citenamefont {Rychetsky}}]{Schranz2020(2)}%
  \BibitemOpen
  \bibfield  {author} {\bibinfo {author} {\bibfnamefont {W.}~\bibnamefont
  {Schranz}}, \bibinfo {author} {\bibfnamefont {A.}~\bibnamefont {Tröster}},\
  and\ \bibinfo {author} {\bibfnamefont {I.}~\bibnamefont {Rychetsky}},\
  }\bibfield  {title} {\bibinfo {title} {Contributions to polarization and
  polarization switching in antiphase boundaries of {SrTiO}$_3$ and
  {PbZrO}$_3$},\ }\href {https://doi.org/10.1063/5.0030038} {\bibfield
  {journal} {\bibinfo  {journal} {Journal of Applied Physics}\ }\textbf
  {\bibinfo {volume} {128}},\ \bibinfo {pages} {194101} (\bibinfo {year}
  {2020}{\natexlab{b}})}\BibitemShut {NoStop}%
\bibitem [{\citenamefont {Ma}\ \emph {et~al.}(2019)\citenamefont {Ma},
  \citenamefont {Fan}, \citenamefont {Tan},\ and\ \citenamefont
  {Zhou}}]{Ma2019}%
  \BibitemOpen
  \bibfield  {author} {\bibinfo {author} {\bibfnamefont {T.}~\bibnamefont
  {Ma}}, \bibinfo {author} {\bibfnamefont {Z.}~\bibnamefont {Fan}}, \bibinfo
  {author} {\bibfnamefont {X.}~\bibnamefont {Tan}},\ and\ \bibinfo {author}
  {\bibfnamefont {L.}~\bibnamefont {Zhou}},\ }\bibfield  {title} {\bibinfo
  {title} {Atomically resolved domain boundary structure in lead
  zirconate-based antiferroelectrics},\ }\href
  {https://doi.org/10.1063/1.5115039} {\bibfield  {journal} {\bibinfo
  {journal} {Applied Physics Letters}\ }\textbf {\bibinfo {volume} {115}},\
  \bibinfo {pages} {122902} (\bibinfo {year} {2019})}\BibitemShut {NoStop}%
\bibitem [{\citenamefont {Fujishita}\ and\ \citenamefont
  {Hoshino}(1984)}]{Fuji1984}%
  \BibitemOpen
  \bibfield  {author} {\bibinfo {author} {\bibfnamefont {H.}~\bibnamefont
  {Fujishita}}\ and\ \bibinfo {author} {\bibfnamefont {S.}~\bibnamefont
  {Hoshino}},\ }\bibfield  {title} {\bibinfo {title} {A study of structural
  phase transitions in antiferroelectric {PbZrO}$_3$ by neutron diffraction},\
  }\href {https://doi.org/10.1143/JPSJ.53.226} {\bibfield  {journal} {\bibinfo
  {journal} {Journal of the Physical Society of Japan}\ }\textbf {\bibinfo
  {volume} {53}},\ \bibinfo {pages} {226} (\bibinfo {year} {1984})},\ \Eprint
  {https://arxiv.org/abs/https://doi.org/10.1143/JPSJ.53.226}
  {https://doi.org/10.1143/JPSJ.53.226} \BibitemShut {NoStop}%
\bibitem [{\citenamefont {Fujishita}\ and\ \citenamefont
  {Katano}(1997)}]{Fuji1997}%
  \BibitemOpen
  \bibfield  {author} {\bibinfo {author} {\bibfnamefont {H.}~\bibnamefont
  {Fujishita}}\ and\ \bibinfo {author} {\bibfnamefont {S.}~\bibnamefont
  {Katano}},\ }\bibfield  {title} {\bibinfo {title} {Re-examination of the
  antiferroelectric structure of {PbZrO}$_3$},\ }\href
  {https://doi.org/10.1143/JPSJ.66.3484} {\bibfield  {journal} {\bibinfo
  {journal} {Journal of the Physical Society of Japan}\ }\textbf {\bibinfo
  {volume} {66}},\ \bibinfo {pages} {3484} (\bibinfo {year} {1997})},\ \Eprint
  {https://arxiv.org/abs/https://doi.org/10.1143/JPSJ.66.3484}
  {https://doi.org/10.1143/JPSJ.66.3484} \BibitemShut {NoStop}%
\bibitem [{\citenamefont {Momma}\ and\ \citenamefont
  {Izumi}(2011)}]{Izumi2011}%
  \BibitemOpen
  \bibfield  {author} {\bibinfo {author} {\bibfnamefont {K.}~\bibnamefont
  {Momma}}\ and\ \bibinfo {author} {\bibfnamefont {F.}~\bibnamefont {Izumi}},\
  }\bibfield  {title} {\bibinfo {title} {{{\it VESTA3} for three-dimensional
  visualization of crystal, volumetric and morphology data}},\ }\href
  {https://doi.org/10.1107/S0021889811038970} {\bibfield  {journal} {\bibinfo
  {journal} {Journal of Applied Crystallography}\ }\textbf {\bibinfo {volume}
  {44}},\ \bibinfo {pages} {1272} (\bibinfo {year} {2011})}\BibitemShut
  {NoStop}%
\bibitem [{\citenamefont {Hlinka}\ \emph {et~al.}(2014)\citenamefont {Hlinka},
  \citenamefont {Ostapchuk}, \citenamefont {Buixaderas}, \citenamefont
  {Kadlec}, \citenamefont {Kuzel}, \citenamefont {Gregora}, \citenamefont
  {Kroupa}, \citenamefont {Savinov}, \citenamefont {Klic}, \citenamefont
  {Drahokoupil}, \citenamefont {Etxebarria},\ and\ \citenamefont
  {Dec}}]{Hlinka2014}%
  \BibitemOpen
  \bibfield  {author} {\bibinfo {author} {\bibfnamefont {J.}~\bibnamefont
  {Hlinka}}, \bibinfo {author} {\bibfnamefont {T.}~\bibnamefont {Ostapchuk}},
  \bibinfo {author} {\bibfnamefont {E.}~\bibnamefont {Buixaderas}}, \bibinfo
  {author} {\bibfnamefont {C.}~\bibnamefont {Kadlec}}, \bibinfo {author}
  {\bibfnamefont {P.}~\bibnamefont {Kuzel}}, \bibinfo {author} {\bibfnamefont
  {I.}~\bibnamefont {Gregora}}, \bibinfo {author} {\bibfnamefont
  {J.}~\bibnamefont {Kroupa}}, \bibinfo {author} {\bibfnamefont
  {M.}~\bibnamefont {Savinov}}, \bibinfo {author} {\bibfnamefont
  {A.}~\bibnamefont {Klic}}, \bibinfo {author} {\bibfnamefont {J.}~\bibnamefont
  {Drahokoupil}}, \bibinfo {author} {\bibfnamefont {I.}~\bibnamefont
  {Etxebarria}},\ and\ \bibinfo {author} {\bibfnamefont {J.}~\bibnamefont
  {Dec}},\ }\bibfield  {title} {\bibinfo {title} {Multiple soft-mode vibrations
  of lead zirconate},\ }\href {https://doi.org/10.1103/PhysRevLett.112.197601}
  {\bibfield  {journal} {\bibinfo  {journal} {Phys. Rev. Lett.}\ }\textbf
  {\bibinfo {volume} {112}},\ \bibinfo {pages} {197601} (\bibinfo {year}
  {2014})}\BibitemShut {NoStop}%
\bibitem [{\citenamefont {\'I\~niguez}\ \emph {et~al.}(2014)\citenamefont
  {\'I\~niguez}, \citenamefont {Stengel}, \citenamefont {Prosandeev},\ and\
  \citenamefont {Bellaiche}}]{Stengel2014}%
  \BibitemOpen
  \bibfield  {author} {\bibinfo {author} {\bibfnamefont {J.}~\bibnamefont
  {\'I\~niguez}}, \bibinfo {author} {\bibfnamefont {M.}~\bibnamefont
  {Stengel}}, \bibinfo {author} {\bibfnamefont {S.}~\bibnamefont
  {Prosandeev}},\ and\ \bibinfo {author} {\bibfnamefont {L.}~\bibnamefont
  {Bellaiche}},\ }\bibfield  {title} {\bibinfo {title} {First-principles study
  of the multimode antiferroelectric transition in {PbZrO}$_3$},\ }\href
  {https://doi.org/10.1103/PhysRevB.90.220103} {\bibfield  {journal} {\bibinfo
  {journal} {Phys. Rev. B}\ }\textbf {\bibinfo {volume} {90}},\ \bibinfo
  {pages} {220103} (\bibinfo {year} {2014})}\BibitemShut {NoStop}%
\bibitem [{\citenamefont {Rabe}(2013)}]{Rabe2013}%
  \BibitemOpen
  \bibfield  {author} {\bibinfo {author} {\bibfnamefont {K.~M.}\ \bibnamefont
  {Rabe}},\ }\bibinfo {title} {Antiferroelectricity in oxides: A
  reexamination},\ in\ \href
  {https://doi.org/https://doi.org/10.1002/9783527654864.ch7} {\emph {\bibinfo
  {booktitle} {Functional Metal Oxides}}}\ (\bibinfo  {publisher} {John Wiley
  \& Sons, Ltd},\ \bibinfo {year} {2013})\ Chap.~\bibinfo {chapter} {7}, pp.\
  \bibinfo {pages} {221--244},\ \Eprint
  {https://arxiv.org/abs/https://onlinelibrary.wiley.com/doi/pdf/10.1002/9783527654864.ch7}
  {https://onlinelibrary.wiley.com/doi/pdf/10.1002/9783527654864.ch7}
  \BibitemShut {NoStop}%
\bibitem [{\citenamefont {Tol\'edano}\ and\ \citenamefont
  {Khalyavin}(2019)}]{Toledano2019}%
  \BibitemOpen
  \bibfield  {author} {\bibinfo {author} {\bibfnamefont {P.}~\bibnamefont
  {Tol\'edano}}\ and\ \bibinfo {author} {\bibfnamefont {D.~D.}\ \bibnamefont
  {Khalyavin}},\ }\bibfield  {title} {\bibinfo {title} {Symmetry-determined
  antiferroelectricity in {PbZrO}$_{3}$, {NaNbO}$_{3}$, and {PbHfO}$_{3}$},\
  }\href {https://doi.org/10.1103/PhysRevB.99.024105} {\bibfield  {journal}
  {\bibinfo  {journal} {Phys. Rev. B}\ }\textbf {\bibinfo {volume} {99}},\
  \bibinfo {pages} {024105} (\bibinfo {year} {2019})}\BibitemShut {NoStop}%
\bibitem [{\citenamefont {Aroyo}\ \emph {et~al.}(2006)\citenamefont {Aroyo},
  \citenamefont {Kirov}, \citenamefont {Capillas}, \citenamefont {Perez-Mato},\
  and\ \citenamefont {Wondratschek}}]{Aroyo2006}%
  \BibitemOpen
  \bibfield  {author} {\bibinfo {author} {\bibfnamefont {M.~I.}\ \bibnamefont
  {Aroyo}}, \bibinfo {author} {\bibfnamefont {A.}~\bibnamefont {Kirov}},
  \bibinfo {author} {\bibfnamefont {C.}~\bibnamefont {Capillas}}, \bibinfo
  {author} {\bibfnamefont {J.~M.}\ \bibnamefont {Perez-Mato}},\ and\ \bibinfo
  {author} {\bibfnamefont {H.}~\bibnamefont {Wondratschek}},\ }\bibfield
  {title} {\bibinfo {title} {{Bilbao Crystallographic Server. II.
  Representations of crystallographic point groups and space groups}},\ }\href
  {https://doi.org/10.1107/S0108767305040286} {\bibfield  {journal} {\bibinfo
  {journal} {Acta Crystallographica Section A}\ }\textbf {\bibinfo {volume}
  {62}},\ \bibinfo {pages} {115} (\bibinfo {year} {2006})}\BibitemShut
  {NoStop}%
\bibitem [{foo()}]{footnote1}%
  \BibitemOpen
  \href@noop {} {}\bibinfo {note} {The translational domain states with respect
  to the oxygen octahedra rotations $\bm \phi$ are not considered in the
  present work.}\BibitemShut {Stop}%
\bibitem [{\citenamefont {Kvasov}\ \emph {et~al.}(2016)\citenamefont {Kvasov},
  \citenamefont {Tagantsev},\ and\ \citenamefont {Setter}}]{Kvasov2016}%
  \BibitemOpen
  \bibfield  {author} {\bibinfo {author} {\bibfnamefont {A.}~\bibnamefont
  {Kvasov}}, \bibinfo {author} {\bibfnamefont {A.~K.}\ \bibnamefont
  {Tagantsev}},\ and\ \bibinfo {author} {\bibfnamefont {N.}~\bibnamefont
  {Setter}},\ }\bibfield  {title} {\bibinfo {title} {Structure and
  pressure-induced ferroelectric phase transition in antiphase domain
  boundaries of strontium titanate from first principles},\ }\href
  {https://doi.org/10.1103/PhysRevB.94.054102} {\bibfield  {journal} {\bibinfo
  {journal} {Phys. Rev. B}\ }\textbf {\bibinfo {volume} {94}},\ \bibinfo
  {pages} {054102} (\bibinfo {year} {2016})}\BibitemShut {NoStop}%
\end{thebibliography}

%\begin{thebibliography}{10}
	
%\end{thebibliography}

\end{document}